\documentclass[useAMS,usenatbib,usegraphicx]{mn2e}

\usepackage{amssymb}
\usepackage{color}
\usepackage{subfig}
\usepackage{fixltx2e}
\makeatletter
\newcommand*{\rom}[1]{\expandafter\@slowromancap\romannumeral #1@}

\newcommand{\zmin}{z_{\rm min}}
\newcommand{\zmax}{z_{\rm max}}
\newcommand{\om}{\Omega_{\rm m}}
\newcommand{\oll}{\Omega_{\Lambda}}

\makeatother

\title[High-redshift standard candles]{High-redshift standard candles: predicted cosmological constraints}
\author[A. L. King et al.]
{\parbox{\textwidth}{Anthea L. King$^{1,2}$\thanks{E-mail:
anthea.king@uqconnect.edu.au}, Tamara M. Davis$^{1}$, K. D. Denney$^{2,3,4}$,  M. Vestergaard$^{2,5}$ and D. Watson$^{2}$ }\vspace{0.4cm}\\
\parbox{\textwidth}{$^{1}$School of Mathematics and Physics, University of Queensland, Brisbane, QLD 4072, Australia\\
$^{2}$Dark Cosmology Centre, Niels Bohr Institute, University of Copenhagen, Juliane Maries Vej 30, DK-2100 Copenhagen, Denmark\\
$^3$ The Ohio State University, Department of Astronomy,140 W. 18th Ave.,Columbus, OH, 43221, USA\\
$^4$ Marie Curie Fellow.\\
$^5$Steward Observatory,
                University of Arizona,
                933 North Cherry Avenue,
                Tucson, AZ 85721. }}

\date{Accepted; Received;}

\pagerange{\pageref{firstpage}--\pageref{lastpage}} \pubyear{2013}

\begin{document}
\label{firstpage}
\maketitle
\begin{abstract}
We investigate whether future measurements of high-redshift standard candles will be a powerful probe of dark energy, when compared to other types of planned dark energy measurements.  Active galactic nuclei, gamma-ray bursts, and certain types of core collapse supernova have been proposed as potential candidates of such a standard candle.  Due to their high luminosity, they can be used to probe unexplored regions in the expansion history of the Universe.  Information from these regions can help constrain the properties of dark energy, and in particular, whether it varies over time.   

We consider both linear and piecewise parameterizations of the dark energy equation of state, $w(z)$, and assess the optimal redshift distribution that a high-redshift standard-candle survey could take to constrain these models.
The more general the form of the dark energy equation of state $w(z)$ being tested, the more useful high-redshift standard candles become.  For a linear parametrization of $w(z)$, they give only small improvements over planned supernova and baryon acoustic oscillation measurements; a wide redshift range with many low redshift points is optimal to constrain this linear model.  However to constrain a general, and thus potentially more informative, form of $w(z)$, having many high-redshift standard candles can significantly improve limits on the nature of dark energy, even compared to dark energy experiments currently only in the planning stages.

\end{abstract}

\section{Introduction}
Cosmological measurements suggest that `dark energy' is the dominant energy component of the Universe, accounting for approximately 70 per cent of the energy density at the present day 
\citep[e.g.][]{Blake2011cos,Blake2011distz,Conley2011,Anderson2012,Hinshaw2012,Padmanabhan2012,Planck2013CosmoParams}. 
 We have no theory that simultaneously explains both its existence and magnitude, which suggests that the standard model of particle physics, quantum physics, or our theory of gravity are incomplete. The simplest model of dark energy corresponds to Einstein's cosmological constant --- a constant energy density, $\rho$, with negative pressure, $p$, such that  $p=-\rho$,
but it could also take a more exotic form such as a dynamical fluid with negative pressure, a scalar potential field, or can be accounted for by a modified theory of gravity such as $f(R)$ \citep{Nojiri2007}. In all models, dark energy can be characterized by its equation of state $w\equiv p/\rho$. Measuring the present value of $w$ and any time variation provides us with crucial information about the underlying physics of dark energy.

The properties of dark energy can be probed by studying its influence on the expansion of the Universe.
Standard candles and rulers are tools for mapping this expansion. Standard candles have had central roles in major cosmological discoveries, from the use of Cepheid variable stars in the discovery of the expanding Universe by \citet{Hubble1929} to the more recent use of Type Ia Supernovae (SNe) in the discovery of the accelerated expansion and  `dark energy' \citep{Riess1998,Perlmutter1999}.

To investigate the Universe and its expansion, it is important to use data sets from various different probes to break degeneracies between cosmological parameters, and therefore derive tighter constraints \citep{Bahcall1999, Huterer2001,Levine2002,  Melchiorri2003,Wang2004, Mantz2010},
Using multiple probes simultaneously is also important for consistency checks between independent methods and to understand and mitigate systematic errors. 
Presently, several cosmological probes are being used to study the Universe, including: SNe, baryon acoustic oscillations (BAO), the cosmic microwave background (CMB), weak lensing (WL), and galaxy clustering (CL).

\begin{figure}
\includegraphics[width=0.5\textwidth]{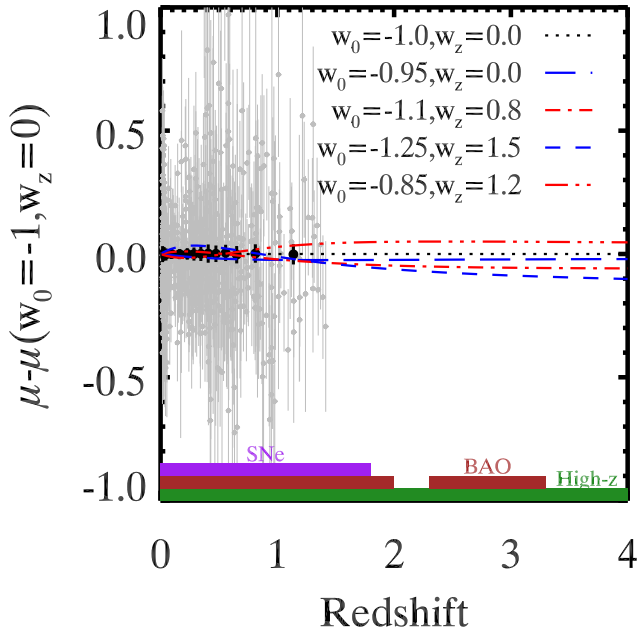}
  \caption{Hubble diagram showing a distance modulus, $\mu$, normalized to that expected for a $\Lambda$CDM universe (dotted black curve). Blue and red curves show possible $w(z)$ models. The models of dark energy we show in this figure are described by the Chevallier-Polarski-Linder, CPL, parametrization, $w(z) = w_0+w_zz/(1+z)$. $\Lambda$CDM is therefore given by $w_0=1$; $w_z=0$.    The Union2 SN data [grey for individual measurements and black for redshift binned results (Weighted arithmetic mean)] are also shown to demonstrate the range of redshifts currently probed by standard candle measurements. All models shown are hard to distinguish with only these data points. The lines with the same colours represent models that are hard to distinguish with only high-redshift measurements (due to uncertainty in the absolute magnitude) but are easy to distinguish with both high- and low-redshift measurements.  The purple, brown, and green shaded regions shown at the bottom of the plot represent the predicted redshift limits of future SN, BAO, and proposed HzSC measurements, respectively. }
  \label{fig:distmods}
\end{figure}

These probes provide us with important and complementary ways to investigate the properties of dark energy, but at present, with the exception of BAO, all of these methods for making cosmological measurements are restricted to relatively low redshifts. Proposed SN measurements may only be observed out to a redshift of $z<2.5$ \citep{Grogin2011,Koekemoer2011}, though the majority of measurements will remain at $z<1.5$, due to their relative faintness at these redshifts and observational magnitude limits, as well as a decrease in SN rates at high redshifts \citep{Albrecht2006, Hook2013}. Galaxy surveys [e.g. \textit{Sloan Digital Sky Survey} (SDSS) \citep{Percival2010}, \textit{WiggleZ} \citep{Blake2011cos}, and \textit{Baryon Oscillation Spectroscopic Survey} (BOSS) \citep{Anderson2012}], from which WL, CL, and BAO measurements are made, are also currently restricted to low redshifts ($z\lesssim 1$). Some BAO measurements  have been obtained in a higher redshift regime by probing distant galaxies through Lyman$-\alpha$ absorption in quasar spectra \citep{Busca2012,Slosar2013}. Future galaxy surveys are predicted to extend the redshift range probed by WL and CL to $z\sim3$ but such surveys will not be completed in the next 5-10 years.
 With information over a larger redshift range, we can more easily identify time-evolving behaviour in dark energy, if it is present. Fig. \ref{fig:distmods} shows that a large range of models are all reasonably consistent with the low-redshift data points and become more easily distinguishable with the inclusion of high-redshift measurements. The models of dark energy we have shown in this figure are described by the Chevallier--Polarski--Linder, CPL, parametrization, $w(z) = w_0+w_zz/(1+z)$.

Active galactic nuclei (AGN) have been proposed as high-redshift cosmological probes by \citet{Watson2011} using a technique called reverberation mapping. AGN display a tight empirical radius-luminosity ($r_{\rm BLR}-L(5100{\rm \AA})$)  relation \citep{Koratkar1991,Wandel1999,Kaspi2000,Bentz2006, Bentz2009,Bentz2013}, which makes them suitable as standardizable candles. Here, $r_{\rm BLR}$ is the distance between the central accretion disc and the broad-emission-line region (BLR), 
where nebular clouds reprocess the accretion disc continuum radiation of luminosity, $L$, into emission line photons. 
 The value of $r_{\rm BLR}$ is measured from the observed time lag between the nuclear continuum and broad-emission-line light-curve variations, taken to be the light travel time between the central source and the BLR \citep[see, e.g.,][]{Peterson2001}. 
 Since AGN are numerous, highly luminous, and persistent sources of light that are present over a broad range of epochs, they are good candidates for distance measurements.   Currently the $r_{\rm BLR}-L$ relationship spans five orders of magnitude in the optical continuum luminosity at 5100\AA~  from $10^{41}$ to $10^{46}$ erg/s with an observed scatter in the relationship as low as  0.13 dex \citep[equivalent to 0.33 mag in the distance modulus; ][]{Bentz2013}, with a clear potential for further reduction in the scatter \citep{Watson2011} making the relationship a reasonable tool for dark energy investigations. 
 
 The $r_{\rm BLR}-L$ relationship is anchored in well-understood photoionization physics \citep{Peterson1997,Osterbrock2006}.
AGN broad emission lines are emitted when photoionization equilibrium is attained in the BLR. For systems with the same ionization parameter, gas densities, and ionizing spectral energy distributions (SEDs), this equilibrium occurs at a specific radius. In AGN, at least to the first order, this condition holds, and as a consequence the simple relationship, $r_{\rm BLR}\propto L^{1/2}$, is expected. This $r_{\rm BLR}-L$ relationship can be translated to $\tau/\sqrt{F}\propto D_L$, where $\tau$ is the measured time delay ($\tau=r_{\rm BLR}/c$), $F$ is the measured flux of the object, and  $D_L$ is the luminosity distance. Thus a Hubble diagram can be constructed - see fig. 2 of \citet{Watson2011}.

 The constancy of the ionization parameter, gas densities, and SED between AGN is supported by the agreement between the predicted and observed $r_{\rm BLR}-L$ relationship \citep{Bentz2009,Bentz2013} and the uniformity in AGN spectra \citep{VandenBerk2001, VandenBerk2004, Dietrich2002}. Despite this, the potential for intrinsic variation in this $r_{\rm BLR}-L$ relationship with black hole characteristics or metallicity, for example,  may need to be tested further.

 
 Besides AGN, gamma-ray burst \citep[GRB][]{Ghirlanda2006,Speirits2007, Liang2008, Diaferio2011,Wei2013}, Type II SNe  \citep[SNe II][]{Poznanski2010}, and the supernovae associated with gamma ray bursts (GRB-SNe, Li \& Hjorth, in prep.)  may also have potential to be standardisable candles, but at this stage, there is no strong evidence to support the use of these probes. New variability surveys make GRBs a highly sought after high-redshift standard candle (HzSC) candidate; however, the physics is still not well known. In contrast, AGN physics is better understood and much of the measurement scatter for AGN can be attributed to known correctable systematics \citep[][Kilerci-Eser et al., in prep.]{Watson2011}.  Accordingly, AGN are likely to be our best candidate at this time. 

In our analysis, we determine the requirements for an HzSC to be a competitive cosmological probe regardless of the type of standard candle. We investigate how this general standard candle can complement existing and future constraints on dark energy properties from Type Ia SNe, BAO, and the CMB. To investigate the properties of dark energy, we consider both a linear, time-evolving dark energy equation of state and a parametrization-independent piecewise equation-of-state model. We also consider how well a general standard candle can measure the Hubble parameter in independent redshift bins, and make an estimate of the dark energy density function. Similar work has been done by \citet{Goliath2001,Huterer2001, Frieman2003, Linder2003,Salzano2013} but only with standard candle measurements with the redshift capabilities of SNe and only in conjunction with CMB measurements. We extend the redshift range of the standard candle probe in this case and also consider the inclusion of BAO constraints. Other authors have looked at how future surveys will constrain dark energy but do not consider the possibility of an HzSC \cite[e.g.,][]{Albrecht2006, Sarkar2008,Eisenstein2011}.

The aim of this paper is to predict the power of standard candle measurements for constraining dark energy properties and determine the optimal redshift distribution required to set the tightest constraints. From this analysis we can make a judgement about how useful HzSCs are as cosmological probes and define an optimal survey strategy. This paper is organized as follows: Section~\ref{sec:prediction} describes the analysis methods we have implemented, Section~\ref{sec:probes} outlines the observables and data sets we use, while Section 4 details the parameters we fit.  We present the results of our analysis in Section~\ref{sec:effect}. Discussion and conclusions are presented in Section~\ref{sec:discuss}.

\section{Analysis Methods}\label{sec:prediction}
We predict the cosmological information that can be extracted from measurements of a reliable HzSC. In particular, we concentrate on how well an HzSC, with an extended redshift range, can further constrain the properties of dark energy over and above existing and predicted future SN, BAO, and CMB measurements.  In order to make these predictions, we employ two different methods: 
\begin{enumerate}
\item Parameter fitting ($\chi^2$), and
\item Fisher matrix analysis.
\end{enumerate}
The parameter fitting method uses a $\chi^2$ analysis on mock standard candle data and tests the likelihood of the data given the model. It is the more accurate of the two methods, but  can be computationally time consuming.  The Fisher matrix method is a second order Gaussian likelihood estimation that is very popular in predicting the constraints on various cosmological parameters due to its simplicity and speed \citep[e.g.][]{Albrecht2006,Bassett2011,More2012}; however, it fails when the errors in the parameter space are non-Gaussian, which is common for individual cosmological probes. Nonetheless, Fisher predictions are reasonably reliable for large numbers of standard candle measurements and for combinations of probes, because as the constraints become tighter, the uncertainties become more Gaussian.

\subsection{Parameter fitting ($\chi^2$)}
Our first method of analysis is likelihood testing using real data from SN, BAO, and CMB measurements and mock catalogues of standard candle data and future SN and BAO data. 
The likelihood that the data are consistent with the model is $\mathcal{L} \propto \exp[-\chi^2/2 ]$.
We explore the parameter space using either a grid approach, for simple models (such as the linear dark energy equation-of-state parametrization), or Markov Chain Monte Carlo (MCMC) analysis, for models with many parameters (such as the piecewise parametrization). We restrict our parameter space to $50.0\leq H_0\leq100.0$ and $0.15\leq\Omega_m\leq0.4$, where $H_0$ is the present day Hubble constant, and $\Omega_m$ is the present matter density fraction.  {Independent measurements of $\Omega_m$ \citep[e.g.][]{Samushia2013} and $H_0$ \citep[e.g.][]{Sandage2006,Riess2011} are consistent with these parameter ranges.  For each trial cosmology in our parameter space, picked by MCMC or grid method, we calculate the $\chi^2$ value,

\begin{equation}
 \chi^2(\mathcal{P}_{\rm mod}) = \sum_{ij} \left[x_i^m(\mathcal{P}_{\rm mod})-x_i^d\right] C_{ij}^{-1}\left[ x_j^m(\mathcal{P}_{\rm mod})-x_j^d\right],
 \label{eq:chi2}
 \end{equation}
 where $x_i^m(\mathcal{P}_{\rm mod})$ is the predicted observable given the model parameters, $x_i^d$ is the observed value,  and $C_{ij}^{-1}$ is the inverse covariance matrix of the observable $x^d$. If the measurements of $x^d$ are independent then the covariance matrix is diagonal, such that $C_{ii} = \sigma_i^2$ where $\sigma_i$ is the uncertainty in the $x_i^d$ measurement.
 
If the observable has the form, $x = f(\mathcal{P})+K$, where $K$ is a constant, and if no prior knowledge of $K$ is assumed at all, we can analytically marginalize over this constant nuisance parameter ($K\in [-\infty,\infty]$). The revised $\chi^2$ equation used for this purpose is given in Appendix \ref{sec:fishermarg}.

\subsection{Fisher Matrices} 
The Fisher matrix method is a  method of predicting constraints on your parameter space without real or simulated data. It is a second order approximation of the likelihood. It assumes Gaussian uncertainty on the parameters being fit, as opposed to only Gaussian uncertainties on the measured quantities as in the previous analysis. Using a fiducial model and expected measurement uncertainties, the likelihood in the nearby parameter space is predicted.

The Fisher matrix is defined such that its inverse is the covariance matrix
\begin{equation}
[\mathcal{F}]^{-1} = [{C}] =  \left[ \begin{array}{cc}
\sigma_{\alpha}^2 & \sigma_{\alpha\beta}            \\
       \sigma_{\alpha\beta} & \sigma_{\beta}^2  \end{array} \right],
       \label{eq:covmatrix}
\end{equation}
where $\sigma_{\alpha}$ is the uncertainty associated with an arbitrary parameter $\lambda_{\alpha}$, and $\sigma_{\alpha\beta} = \rho\sigma_{\alpha}\sigma_{\beta}$ where $\rho$ is known as the correlation coefficient, which varies from 0 (independent) to 1 (completely correlated). Once the Fisher matrix is known, the inverse gives the best possible constraints that we can derive for the parameters given the observed data. According to the Cramer-Rao inequality, the Fisher matrix gives a lower bound on the parameter uncertainty $\sigma$ in parameter $\lambda_{\alpha}$,
\begin{equation}
\sigma_{{\alpha}} \geq \sqrt{\left(F^{-1}\right)_{\alpha\alpha}}.
\end{equation}

For $N$ model parameters $\lambda_{\alpha},\lambda_{\beta},\ldots, \lambda_N$, the Fisher matrix, $F$, is an $N\times N$ symmetric matrix.  Consider $b$ observables, $f_1,f_2,\ldots,f_b$ (such as  $\mu$), with which you attempt to measure cosmological parameters, $\lambda_n$  (such as $\Omega_m, \Omega_{\Lambda},  w$), where each observable is related to the model parameters by some function, e.g. $\mu = \mu(\Omega_m,\Omega_{\Lambda}, \ldots)$. Then the elements of the Fisher matrix are given by,
\begin{equation}
F_{\alpha\beta} = \sum_b \frac{1}{\sigma_{b}^2} \frac{\partial f_b}{\partial \lambda_{\alpha}}\frac{\partial f_b}{\partial \lambda_{\beta}},
\label{eq:genfisher}
\end{equation}
where each element is summed over the observables. The derivatives of the parameters required for the Fisher matrix analysis are given in Appendix~\ref{sec:fishercalc}. 

To marginalize over any parameter, the Fisher matrix is inverted, then the associated rows and columns for that parameter are removed from the matrix, and the inverted Fisher matrix is once again inverted to give the revised Fisher matrix.
Useful formulae for performing this marginalization stably are described in the appendix of \citet{Matsubara2004}. 

Computationally Fisher matrices are much simpler than performing the full likelihood analysis and is therefore common practice when forecasting the precision of a future survey \citep{Albrecht2006, Eisenstein2011}. 

\subsection{Quality Measures of Constraining Power}
To quantify the improvement achieved with the addition of a new HzSC, we calculate the predicted change in the constraints of various dark energy parameters.  These include the Hubble parameter in several redshift bins $H(z_i)$, a piecewise fit to the dark energy equation-of-state $w(z_i)$, and a linear parametrization of a time-varying dark energy equation-of-state, $w(z)=w_0+w_zz/(1+z)$ . 
In the latter case, we also consider the figure of merit (FoM) suggested by the Dark Energy Task Force  \citep[DETF; ][]{Albrecht2006},  given by the inverse of the area within the 95 per cent confidence level ($2\sigma$) contours of the parameters $w_0$ and $w_z$,
\begin{equation}
{\rm FoM} = \frac{1}{\Delta\chi^2\pi\sqrt{\det {\rm Cov}(w_0,w_z)}},
\label{eq:FoMproper}
\end{equation}
where $\Delta\chi^2 = 6.17$ for two parameters.\footnote{In general, if the likelihood surfaces for all the parameters are Gaussian, any $N$-dimensional volume is proportional to the square root of the determinant of the covariance matrix of $\{\lambda_i\}$, $\sqrt{\det {\rm Cov}(\lambda_1,\lambda_2,\ldots)}$. For $N=2$, the 1$\sigma$ or 2$\sigma$ confidence level contours of the parameters $\lambda_1$ and $\lambda_2$ are ellipses with the enclosed area given by $\Delta\chi^2\pi\sqrt{\det {\rm Cov}(\lambda_1,\lambda_2)}$, where $\Delta\chi^2$ given by 2.30 or 6.17, respectively \citep{Wang2008}. }  Despite equation \ref{eq:FoMproper} being the definition of the FoM that appears in the DETF report, the more recognizable form of the DETF FoM is given by the expression $[\sigma~w_p\times\sigma~w_a]^{-1}$, which is equivalent to
\begin{equation}
[\sigma~w_p\times\sigma~w_a]^{-1}= \frac{1}{\sqrt{\det {\rm Cov}(w_0,w_z)}} \approx 19.38\times {\rm FoM},
\end{equation}
where $w_p=w(z_p)$ is the dark energy equation-of-state value at the pivot redshift, and  the pivot redshift is the redshift at which $w(z)$ has the smallest uncertainty. The transformation between ($w_0,~w_z$) coordinates to ($w_p,~w_a$) coordinates is linear  and the Fisher matrix in the ($w_p,~w_a$) variables is $F'=M^TFM$, with $\det M=1$. It follows that the error ellipse in the $w_p- w_a$ plane has the same area as the equivalent ellipse in the $w_0- w_z$ plane.

We show both values of FoM in our results for simplicity. Throughout our analysis, we assume a flat universe. As a consequence, our results are not directly comparable with those from the DETF, who allowed for curvature. We  only consider a flat universe due to the strength of the current constraints on the curvature given by CMB measurements \citep{Planck2013}.

\section{Observables and Data}\label{sec:probes}
Standard candle measurements provide us with the luminosity distance, $D_L$, through the relationship between measured flux, $F$, and intrinsic luminosity. Because $F$ is measured without an absolute luminosity calibration, what is actually calculated from the observables is the distance modulus,
\begin{equation}
\mu = m-M = 5 \log[D_L(\mathcal{P},z)] + \mathcal{M}
\label{eq:distmod}
\end{equation}
where $\mathcal{M} = \log_{10} (c/H_0) +25$, is a constant over which we marginalize (equation \ref{eq:revisechi2} absorbs the uncertainty in the absolute magnitude, $M$, as well as the uncertainty in $H_0$), and $\mathcal{P}$ are the parameters that describe the universe and influence the expansion. The luminosity distance, $D_L$, is given by,
\begin{equation}
D_L = (1+z)D_M
\end{equation}
where $z$ is the redshift and $D_M$ is the comoving tangential distance, defined as
\begin{equation}
D_M = \frac{c}{H_0} \frac{1}{\sqrt{\Omega_k}} \sinh(\sqrt{\Omega_k}\chi)  = R_0  \sinh(\sqrt{\Omega_k}\chi),
\end{equation}
where $R_0 = c/(H_0 \sqrt{\Omega_k})$ is the present day scale factor, and $\Omega_k$ is the equivalent energy density fraction of the curvature.
The dimensionless comoving distance is,
\begin{equation}
\chi(z) =  \int_0^z \frac{H_0}{H(z)} dz 
\end{equation}

The general form of the Hubble parameter, $H(z)$,  for a Friedman-Robertson-Walker metric is given by
\begin{eqnarray}
\frac{H(z)^2}{H_0^2} = \Omega_r(1+z)^4 + {\Omega_m}(1+z)^3 + {\Omega_k}(1+z)^2  \nonumber\\ + {\Omega_x}\exp\left\{3\int_0^{1/(1+z)} \frac{dz}{1+z}\left[1+w(z)\right]\right\} ,
\label{eq:Hubblegeneral}
\end{eqnarray}
where $\Omega_r$ is the current normalized radiation energy density (including the relativistic neutrino density),  and $\Omega_x$ is the current normalized dark energy density. The  energy density is  normalized with the critical density, such that $\Omega=\rho/\rho_c$, where $\rho_c$ is the critical energy density of the universe for which the spatial geometry is flat (or Euclidean).
\subsection{Data}\label{sec:data}
\subsubsection{Existing Data}
\paragraph*{Type Ia supernova:}
 For our analysis we use the Supernova data from the \textit{SuperNova Legacy Survey} \citep[SNLS;][]{Conley2011} which is a compilation of the first three-year results from the SNLS survey with other supernova surveys (Contains: 123 low-$z$, 93 SDSS, 242 SNLS, and 14 \textit{Hubble Space Telescope} SN measurements). The details of our fitting procedure, including stretch, and colour corrections, are discussed in Appendix~\ref{sec:Snestuff}.
\paragraph*{Baryon acoustic oscillations:}
Large-scale structure measurements and consequently BAO measurements can be distilled into simple parameters that can be used to constrain cosmology. Two common parameters are used to express the cosmological information from the BAO measurement:  $A(z)$ and $d_z$.  The acoustic parameter, $A(z)$, was introduced by \citet{Eisenstein2005} and  is given by
 \begin{equation}
A(z) = \frac{D_V(z)\sqrt{\Omega_m H_0^2}}{cz},
\end{equation}
where the $D_V$ is the `dilation scale' distance,
\begin{equation}
D_V(z) = \left(D_M^2\frac{cz}{H(z)}\right)^{1/3}.
\end{equation}
The ratio of the sound horizon scale to the dilation scale was given the symbol $d_z$ by \citet{Percival2010},
\begin{equation}
d_z = \frac{r_s(z_d)}{D_V(z)},
\end{equation}
where $z_d$ is the redshift at the `baryon-drag epoch' and $r_s(z_d)$ is the comoving sound horizon at the baryon-drag epoch. The general expression for $r_s(z)$ is given by \citep{Komatsu2011}
\begin{equation}
r_s(z) = \frac{c}{\sqrt{3}}\int_0^{1/(1+z_d)} \frac{da}{a^2H(a)\sqrt{1+(3\Omega_b/4\Omega_{\gamma})a}}
\end{equation}
where $\Omega_{\gamma} = 2.469\times10^{-5}(T_{\rm eff}/2.725)^4$ is the normalized pure radiation density, $a$ is the normalized scale factor and is related to the redshift by $a^{-1}=1+z$, and  $T_{\rm eff}$ is the effective temperature of the CMB.  We use the fitting formula for $z_d$ defined by \citet{EisensteinHu1998}.

The values of these two parameters from \textit{Six-degree-field Galaxy Survey} (6dFGS) \citep{Beutler2011}, SDSS \citep{Percival2010}, WiggleZ \citep{Blake2011distz}, and BOSS \citep{Anderson2012} are shown in Table~\ref{tab:acousticparameter}.
\begin{table}
  \caption{The BAO distance data set from the 6dFGS, SDSS, WiggleZ and BOSS surveys.  }
    \begin{tabular}{cccc}
    \hline
   Survey & $z$     & $d_z$ &$ A(z)$ \\
    \hline
    6dFGS & 0.106 &  $\mathbf{0.336 \pm 0.015}$ & $0.526 \pm 0.028$ \\
    SDSS  & 0.2   & $\mathbf{0.1905 \pm 0.0061}$ & $0.488 \pm 0.016$ \\
    SDSS  & 0.35  &  $\mathbf{0.1097 \pm 0.0036}$ & $0.484 \pm 0.016$ \\
    WiggleZ & 0.44  & $0.0916 \pm 0.0071$ &  $\mathbf{0.474 \pm 0.034}$ \\
    WiggleZ & 0.6   & $0.0726 \pm 0.0034$ & $\mathbf{0.442 \pm 0.020}$ \\
    WiggleZ & 0.73  & $0.0592 \pm 0.0032$ & $\mathbf{0.424 \pm 0.021}$ \\
    BOSS & 0.57& $\mathbf{0.0731\pm0.0018}$ &-\\
    \hline \\
    \end{tabular}%
    \medskip \\
      Notes: {Measurements of the distilled parameters $d_z$ and $A(z)$ are quoted. The values in \textbf{bold} are the values we have used in our analysis.}
  \label{tab:acousticparameter}%
\end{table}%
We use the $d_z$ parameter for our analysis of the 6dfGS, SDSS and BOSS data and the $A(z)$ parameter when we are considering the WiggleZ data, corresponding to the officially released parameters in the associated papers. These parameters provide the best depiction of the BAO data in each survey. The distilled parameter, $A(z)$, is the most appropriate choice for quantifying the WiggleZ data as it is uncorrelated with $\Omega_mh^2$, but because of the shape of the clustering pattern marginalized over for the 2dfGS, SDSS, and BOSS data, the $d_z$ parameter is the best representation.
Therefore, for the $\chi^2$ analysis we define $x^d= \left[d_{0.106} [{\rm 6dfGS}]\right.$, $d_{0.2}[{\rm SDSS}]$, $d_{0.35} [{\rm SDSS}]$, $A(0.44) [{\rm WiggleZ}]$, $A(0.6) [{\rm WiggleZ}]$, $A(0.73) [{\rm WiggleZ}]$, $d_{0.57} [{\rm BOSS}]\left.\right]$ and $C_{ij}^{-1}$ is a $7\times7$ matrix made up from a combination of the individual errors from the individual measurements from 6dfGS and BOSS and the defined covariance matrices from the WiggleZ \citep{Blake2011distz} and SDSS    \citep{Percival2010} data. We have not included any covariance terms between the surveys, despite the fact that the WiggleZ and SDSS surveys share a sky overlap of 500 square degree for redshift range $z < 0.5$. Given that the SDSS measurement is derived across an 8000 square degree sky area and the uncertainties in both measurements contain a significant shot noise component, the resulting covariance is negligible \citep{Blake2011distz}. We have assumed Gaussian errors in the BAO distances. Non-Gaussian tails may be non-negligible \citep{Percival2007, Percival2010, Bassett2010}, but studying their effect is beyond the scope of this paper.

The parameters we consider in the cosmological fitting are $\lambda_{\alpha}=[\Omega_m,~w,~H_0,~\Omega_b]$. We marginalize over $\Omega_b$ and $H_0$ values  after the various probes are combined as the cosmological parameters do not have independent probability distributions.
\paragraph*{Cosmic microwave background (CMB):}
We included the CMB data in our cosmological fits using the \textit{Planck}
 \citep{Planck2013CosmoParams} results.  We use the CMB `distance priors': the shift parameter, $\mathcal{R}$, the acoustic parameter, $\ell_A$, and the  redshift at the decoupling epoch,  $z_{*}$. The shift parameter is given by 
\begin{equation}
\mathcal{R} = \frac{\sqrt{\Omega_m H_0^2}}{c}D_M(z_{*}). 
\end{equation}
The acoustic parameter is given by the expression
\begin{equation}
\ell_A = \frac{\pi D_M(z_{*})}{r_s(z_{*})} 
\end{equation}
where $r_s(z_{*})$ is the sound horizon at recombination. The redshift at photon decoupling is  $z_{*}$ and we implement the \citet{HuSugiyama1996} fitting formula. 

The measured \textit{Planck} `distance priors' (D. Parkinson 2013, private communication) from the \citet{Planck2013CosmoParams} are $[\mathcal{R},\ell_A ,z_{*}]= [1.7440\pm0.011, 301.62\pm0.19,1090.02\pm0.42]$. These parameters capture most of the constraining power of the CMB data for the dark energy properties \citep{Komatsu2009}. \citet{WangWang2013} published values for $\mathcal{R}$ and $\ell_A$ from the Planck data but did not include a value of $z_{*}$. Their findings are consistent with the values stated here.

We again consider the parameters $\lambda_{\alpha}=[\Omega_m,~w,~H_0,~\Omega_b]$,
 marginalizing over $\Omega_b$ and $H_0$ values  after the various probes are combined. 

\subsubsection{Mock data}
Mock catalogues were constructed to simulate future 
 HzSC measurements (AGN, GRB, SN II) and future SN and BAO measurements. The future SN and BAO predictions are taken from the Stage III and IV predictions from \citet{Albrecht2006}. Stage III are intermediate-scale, near-future projects, and Stage IV are large-scale, longer-term future projects. Table \ref{tab:stage3and4surveys} shows proposed Stage III and IV surveys.

Our fiducial model is set as a flat $\Lambda$ cold dark matter ($\Lambda$CDM) universe with the maximum likelihood parameters we determined from the joint SNLS3, SDSS, 6dF, WiggleZ, BOSS, and \textit{Planck} constraints, using an MCMC chain with $w=-1$:
 \begin{description}
 \item [$\Omega_m = 0.30$] (matter energy density), 
 \item [$H_0 = 69.45$] (the Hubble constant),
 \item [$N_{\rm eff} = 3.04$] (effective number of neutrino-like relativistic degrees of freedom),
 \item [$T_{\rm eff} = 2.7255$] (effective temperature).
 \end{description}


 \begin{table*}
  \caption{Proposed future Stage III and IV cosmological surveys  \citep{Yoo2012}. Dark energy projects are classified into four stages:
Stage I-completed projects that have already released data, Stage II-on-going projects, Stage
III-intermediate-scale, near-future projects, and Stage IV-large-scale, longer-term future
projects. }
    \begin{tabular}{rp{3cm}p{3cm}p{3cm}p{3cm}}
    \hline
    Probes & SN Ia & CMB   & BAO   & WL \\
    \hline
    Stage III & DES, Pan-STARRS4, ALPACA, ODI   & ALPACA, CCAT  & DES, HETDEX, BigBOSS, ALPACA  & DES, Pan-STARRS4, ALPACA, ODI \\
    Stage IV & LSST, WFIRST, SNAP,  JWST & EPIC, LiteBIRD, B-Pol & LSST, SKA, WFIRST, Euclid, JWST & LSST, SKA, WFIRST, Euclid \\
    \hline
    \end{tabular}\\
    \raggedright
    \underline{References:}\\
    ALPACA \citep{Corasaniti2006}, BigBOSS \citep{BigBOSS}, B-Pol \citep{DeBernardis2009}, CCAT \citep{Radford2007}, DES \citep{Lin2006},  EPIC \citep{Bock2009}, Euclid \citep{Laureijs2011},  HETDEX \citep{Hill2008}, JWST \citep{JWST2009}, LiteBIRD \citep{Hazumi2011},  LSST \citep{Ivezic2008}, ODI \citep{ODI2002}, Pan-STARRS4 \citep{PanSTARRS2004}, SKA \citep{TorresRodriguez2007}, SNAP \citep{SNAP2007}, WFIRST \citep{Spergel2013}.
  \label{tab:stage3and4surveys}%
\end{table*}%

\paragraph*{Mock high-$z$ standard candle measurements}
 a standard candle mock catalogue is constructed by generating perfect distance modulus data according to our fiducial model, and adding random Gaussian error of the order of the predicted scatter. 
 
 We generate various mock catalogues for a number of mock surveys varying the redshift range and distribution. We assume, unless specifically mentioned, that the uncertainty in the distance modulus measurement for the standard candle is 0.2 mag, chosen following the expected achievable scatter discussed by  \citet{Watson2011} for AGN measurements. We generally consider a large mock standard candle catalogue with 2000 distance measurements. This number was chosen as it is directly comparable to Stage III SN numbers for the individual predicted spectroscopic or photometric surveys. We extend this study to consider AGN distributions and realistic observational restrictions in an upcoming paper.

We have assumed independence between individual standard candle measurements. Correlations could be induced by shared peculiar velocities if close enough to be influenced by the same overdensity (e.g. galaxy cluster) or by lensing magnification if close to the same line of sight. However our HzSC candidates are typically at a high enough redshift that peculiar velocity effects are negligible and widely spread enough over the sky that nearby lines of sight are rare, therefore assuming the individual measurements are not correlated is reasonable. As a consequence, the covariance matrix $C_{ij}$ is a diagonal matrix where $C_{ii} = \sigma_{\mu_i}^2$.

\paragraph*{Future SN and BAO constraints}
the mock future SN and BAO measurements are also constructed according to our fiducial model with Gaussian scatter. The future BAO and SN predictions are taken from the Stage III and IV
 predictions from \citet{Albrecht2006}. 
 
 The predicted SN measurements from DETF \citep{Albrecht2006},  are limited to $z<1.7$, but future surveys on the \textit{Hubble Space Telescope} and \textit{James Webb Space Telescope} (JWST) are now expected to observe SNe  out to $z<2.5$ \citep{Grogin2011,Koekemoer2011} and possibly $z<3.5$ \citep{Hook2013}. For consistency with the DETF predictions we do not include these objects in our future SN predictions.  The small number of objects they will find should be considered as part of our predictions for HzSCs.  \citet{Salzano2013} have investigated how the high-$z$ SN measurements improve existing SN constraints.

The predicted measurements given by DETF \citep{Albrecht2006} break the BAO measurement into the perpendicular and transverse components rather than the previously described angle-averaged measurements ($d_z$ and $A(z)$). 
The predicted DETF BAO constraints are consequently given in terms of $\log(D_M(z))$, and $\log(H(z))$. We follow this prescription in our future BAO predictions. We have used both the ground and space based predictions for the BAO and SN predictions. The specifications of the predicted Stage III and IV surveys are described in Table \ref{tab:stage3and4specs}.
We have chosen an intermediate value for the systematic error values for both the SN and BAO measurements [$\sigma_F=0.03$ (associated with photo-$z$ uncertainty),~$\sigma_{L/Q}=0.02$ (associated to the linear and quadratic components of $z$ dependent SN evolution), see \citet{Albrecht2006} for full description] such that the resulting predictions lie directly between the optimistic and pessimistic cases.

\begin{table*}
\caption{The specifications of the predicted Stage III and IV SN and BAO survey measurements.}
\begin{center}
\begin{tabular}{cccccccc}
\hline
SNe	 & &   & &  & & & \\ 
\hline
  Stage &	Type &	N &	Redshift Range &	${\sigma_D}^{a}$ & 	${\sigma_F}^{b}$ &	${\sigma_{L/Q}}^{c}$& Expected year of Completion\\ 
  III &	Spectroscopic &	2001 &	$0.01<z<1.0$ &	0.15 &	0.00 &	0.02/$\sqrt{2}$& 2017 (HETDEX) \\ 
  III &	Photometric &	2001	 &$0.01<z<1.0$ &	0.12 &	0.03 &	0.02/$\sqrt{2}$& 2017 (DES)\\ 
  IV &	Spectroscopic &	2498 &	$0.01<z<1.7$ &	0.10 &	0.00 &	0.02/$\sqrt{2}$& $>2020$ (SNAP, WFIRST),  2023 (JWST)\\ 
 IV &	Photometric &	191679 &	$0.01<z<1.2$ &	0.10 &	0.03&0.02/$\sqrt{2}$& 2032 (LSST)\\ 
\hline
  BAO &  &  & &   &   &   \\ 
\hline
  Stage &	Type	 &Sky Area (deg$^2$) &	Redshift Range &	${\sigma_F}^{b}$&& & Expected year of Completion  \\ 
  III &	Spectroscopic &	2000	 &$0.5<z<1.3$ &	0.00& & & 2014 (BOSS), 2017 (HETDEX) \\ 
  III &	Spectroscopic &	300 &	$2.3<z<3.3$ &	0.00 &&  & 2014 (BOSS)   \\ 
  III &	Photometric &	4000 &	$0.5<z<1.4$ &	0.03&& &   2017 (DES)   \\ 
  IV &	Spectroscopic &	20000 &	$0.01<z<1.5$ &	0.00& &  & 2021 (BigBOSS), $> 2024$ (SKA)    \\ 
  IV &	Spectroscopic &	10000 &	$0.5<z<2.0$ &	0.00  & & &2021 (JDEM), 2023 (JWST) \\ 
  IV &	Photometric &	20000 &	$0.2<z<3.5$ &	0.03 &&   & 2032 (LSST)  \\ 
\hline
\end{tabular}
\end{center}
\raggedright
$^{a}$ The uncertainty of the corrected apparent magnitudes due solely to variations in the properties of SNe.\\
$^{b}$ The uncertainty associated with photometrically determined redshifts.\\
$^{c}$ The uncertainty associated with any redshift dependence in the SN population. The $L$ and $Q$ stand for the linear and quadratic components of evolution.\\
\underline{References:}\\
HETDEX \citep{HETDEXdate}, DES  \citep{DESdate}, SNAP/JDEM \citep{JDEM2009}, WFIRST \citep{DecadalPlan2010}\, LSST  \citep{LSSTdate}, BOSS \citep{BOSSdate}, BigBOSS \citep{BigBOSSdate}, JWST \citep{JWSTdate}.
\label{tab:stage3and4specs}
\end{table*}%

We did not consider future CMB constraints at this point as no survey is predicted to supersede Planck, nor will we consider the constraints from WL and clustering measurements, both from redshift surveys and X-ray identification, as it is beyond the scope of this study. 

\section{Fitting Parameters}
\subsection{Hubble Parameter determination}\label{sec:Hubble}
For a flat universe, by transforming the distance modulus into a comoving distance we can extract an estimate of the Hubble parameter at $z$ through numerical differentiation as
\begin{equation}
H(z) = \frac{1}{c} \left[\frac{{\rm d}D_M(z)}{{\rm d}z}\right]^{-1}.
\label{eq:piecewisehubble}
\end{equation}
This technique was proposed by \citet{Wang2005} and allows an independent determination of the Hubble parameter at different redshifts. The Hubble parameter can also be measured through various other techniques, such as BAO measurements and age--redshift relationships.  Table~\ref{tab:hubble} in Appendix \ref{sec:HubbleParamMeas} shows existing measurements of the Hubble parameter and their associated techniques. For a generic dark energy density evolutions, $\rho_x(z)$, the general Hubble parameter formulation given in Equation \ref{eq:Hubblegeneral} can be simplified to take the form,
\begin{eqnarray}
\frac{H(z)^2}{H_0^2} = \Omega_r(1+z)^4 + {\Omega_m}(1+z)^3 + {\Omega_k}(1+z)^2 +  {\Omega_x}\frac{\rho_x(z)}{\rho_x(0)}.\nonumber \\
{}
\end{eqnarray}
Given precise measurements of the current matter density fraction $\Omega_m$, and assuming a flat universe with a relatively negligible current radiation density fraction $\Omega_r$, the dark energy density function, $\rho_x(z)/\rho_x(0)$, can trivially be determined from $H(z)$ at low redshifts. Here, $\rho_x(0)$ is the current dark energy density fraction.  We set $\Omega_m=0.261\pm0.037$ as determined from anisotropic clustering of galaxies in the CMASS DR9 \citep{Samushia2013} in combination with $H_0$ measurements \citep{Riess2011}, and  the full \textit{Wilkinson Microwave Anisotropy Probe 9} (WMAP9) likelihood \citep{Hinshaw2012} for a $w_z$CDM model. For our analysis we broke the SN and HzSC measurements into evenly spaced redshift bins, to match the convention of the Stage III and IV predictions.

\subsection{Dark energy equation-of-state: $w(z)$}
All models of dark energy can be characterized by their equation-of-state $w$, which may evolve with time. Therefore, crucial information about the underlying physics of dark energy can be obtained by measuring the present value of $w$ and any time variation. The two main strategies for investigating the evolution of the dark energy equation-of-state are to (i) assume a $w(z)$ parametrization and fit to existing data or (ii) determine the value of $w(z)$ in different redshift bins, independent of a dark energy parametrization. The first method can more precisely determine the $w(z)$ behaviour if the parametrization represents the true dark energy evolution. If the dark energy behaves differently than the predicted parametrization, this approach is  possibly misleading. The second approach is statistically noisy as it depends on the first and second derivatives of the distance with respect to redshift. A consequence is that the uncertainties on the measurements of $w(z)$ can become substantial.  On the other hand, it does not require any a priori assumptions about the properties of the dark energy and, as such, can more easily identify exotic behaviour.  We will consider both approaches in our analysis.

\subsubsection{Linear $w(z)$ parametrization}\label{sec:parameter}
We initially consider the linear parametrization of the dark energy equation of state given by the expression,
\[w(z)=w_0+w_z z/(1+z).\]
This parametrization was first proposed by \citet{Chevallier2001} and \citet{LinderCPL2003} and is commonly used throughout the literature\footnote{This parametrization is equivalent to the common $w(a)= w_0+(1-a)w_a$, where $w_z=w_a$ but we use the notation $w_z$ as we primarily refer only to redshift in our analysis.}. For a cosmological constant ($\Lambda$CDM) model, the dark energy equation of state is characterized by $w_0=-1$ and $w_z=0$. The dimensionless Hubble parameter for this parametrization is given by,
\begin{eqnarray}
\frac{H(z)^2}{H_0^2} &=& \Omega_r(1+z)^4 +  {\Omega_m}(1+z)^3 + {\Omega_k}(1+z)^2  \nonumber\\
&& +  \;{\Omega_x}(1+z)^{3(1+w_0+w_z)} e^{-3w_z z/(1+z)}.
\end{eqnarray}

\subsubsection{Redshift-binned piecewise $w(z)$}\label{sec:wzpiece}
Next we consider the value of $w(z)$ in different redshift bins, independent of a dark energy parametrization.  Under the assumption of general relativity, the dark energy equation of state can be expressed as 
\begin{equation}
w(z) = -\frac{(2/3)(1+z)(\frac{{\rm d}\chi}{{\rm d}z})^{-1}\frac{{\rm d}^2\chi}{{\rm d}z^2}-1}{1-(\frac{{\rm d}\chi}{{\rm d}z})^{2}\Omega_m (1+z)^3},
\label{eq:wzdirect}
\end{equation}
where $\chi$ is the dimensionless comoving distance defined earlier \citep{Daly2004}. Determining this directly in independent redshift bins, through numerical differentiation, as was done for the Hubble parameter in Equation \ref{eq:piecewisehubble}, is very noisy due to data limitations and the discreteness of the individual measurements.  Instead, we consider $w(z)$ as a piecewise function, with a constant equation-of-state parameter within each redshift bin, and fit the parameters $w_1,w_2,\ldots, w_i$ using the Monte Carlo analysis, where $w_i$ is the dark energy equation of state corresponding to the $i$th redshift bin, $z_i$. No priors are put on the value of $w_i$.

By choosing $w(z)$ to be a  piecewise constant function (or step function), rather than calculating $w(z)$ directly in each redshift bin as in Equation~\ref{eq:wzdirect}, we can fit all the data at once using MCMC and easily incorporate existing SN, BAO, and CMB data, and mock catalogues of future data. This maximizes the information that can be gleaned by the finite number of distance measurements in our samples. However, this process creates correlations in $w(z)$ between the bins, as the distance, $D_M(z)$, requires an integration over 0 to $z$. The correlations are captured by the covariance matrix, given by $[C] = \left\langle {w} {w^T} \right\rangle -\left\langle {w}\right\rangle\left\langle {w^T}\right\rangle $. To decorrelate the equation-of-state parameters, we follow the prescription of \citet{Huterer2005} and transform the $w$ chains through an orthogonal matrix rotation that diagonalizes the inverse covariance matrix. This is equivalent to applying a weighting function to the correlated $w_i$ values. The new, uncorrelated $w_i$ are given as a linear combination of the correlated $w_i$ described by the weight function. 

The dimensionless Hubble parameter in this case is given by the expression
\begin{eqnarray}
\frac{H(z_{N-1}< z \leq  z_N)^2}{H_0^2} = \Omega_r(1+z)^4 + {\Omega_m}(1+z)^3 + {\Omega_k}\nonumber\\ \times
(1+z)^2 +  {\Omega_x}\left({1+z}\right)^{3(1+w_N)}\prod_{i=0}^{N-1} \left[{1+{\rm max}(z_i)}\right]^{3(w_i-w_{i+1})}
\end{eqnarray}
where $N$ is the redshift bin where $z$ resides and max$(z_i)$ is the maximum redshift in the $i$th redshift bin.

 We divide the redshift range into the following five bins: $0.0\leq z_1<0.3$, $0.3\leq z_2<0.8$, $0.8\leq z_3<1.2$, $1.2\leq z_4<4.0$, $4.0\leq z_5$.  The first four bins are constrained by SN, BAO, and HzSC measurements (roughly a low redshift bin, two medium redshift bins, and a high redshift bin), and the highest redshift bin is constrained entirely by CMB measurements, and therefore is largely uncorrelated with the lower redshift bins (i.e., there is no contribution from preceding bins in the weighting function of bin 5).  We assume a flat universe and evaluate $w_i$ in each redshift bin $i$ by fitting the data. 
 
\section{Results: Constraints from High-Redshift Standard Candles}\label{sec:effect}
In this section, we quantify the power of standard candle measurements for constraining dark energy properties using the analysis methods described above.  Our primary concern is the optimal redshift range for future standard candle measurements. 

Defining the  optimal redshift range is difficult, as it will depend on the model of $w$ being tested. The redshift range that best constrains a constant $w$ will not be the same as that which best constrains a variable $w$. In this section we investigate this multidimensional question and discuss several aspects of the result.

Realistically, the number density of standard candles and the observing capabilities and strategy of a survey will set the number, the measurement uncertainties, and the redshift distribution of standard candle measurements. We consider the effect of observational restrictions on our cosmological predictions, for the specific case of AGN as our standard candle in a  forthcoming paper. For the time being, we consider uniform and non-uniform distributions of a general standard candle, spanning a range of redshifts.

For most of our analysis, we keep the number and scatter in our mock HzSC measurements fixed and alter only the redshift distribution. Greater numbers and higher precision will both give monotonic advantages in constraining dark energy. This is obvious from the role of $\sigma$ in Equations (\ref{eq:chi2}) and (\ref{eq:genfisher}). If we simply assume the main sources of scatter in the standard candle are observational (statistical), rather than intrinsic (systematic), we can consider the improvement in parameter constraints as a trade-off between the number of measurements and the precision of the measurements.  The resulting constraints follow the general relationship $\sigma^2_{\alpha} \propto \sigma_\mu^2/N$, where ${\alpha}$ can represent ($\Omega_m,~\Omega_x,~w,~\ldots$) and $\sigma_\mu$ is the uncertainty in the distance modulus.
 We illustrate this in Fig.~\ref{fig:sigmanumplot} for a flat $w_z$CDM parametrization and an optimal double Gaussian distribution of standard candle measurements, which we discuss in the next subsection. 

\begin{figure*}
		\includegraphics[width=1.0\textwidth]{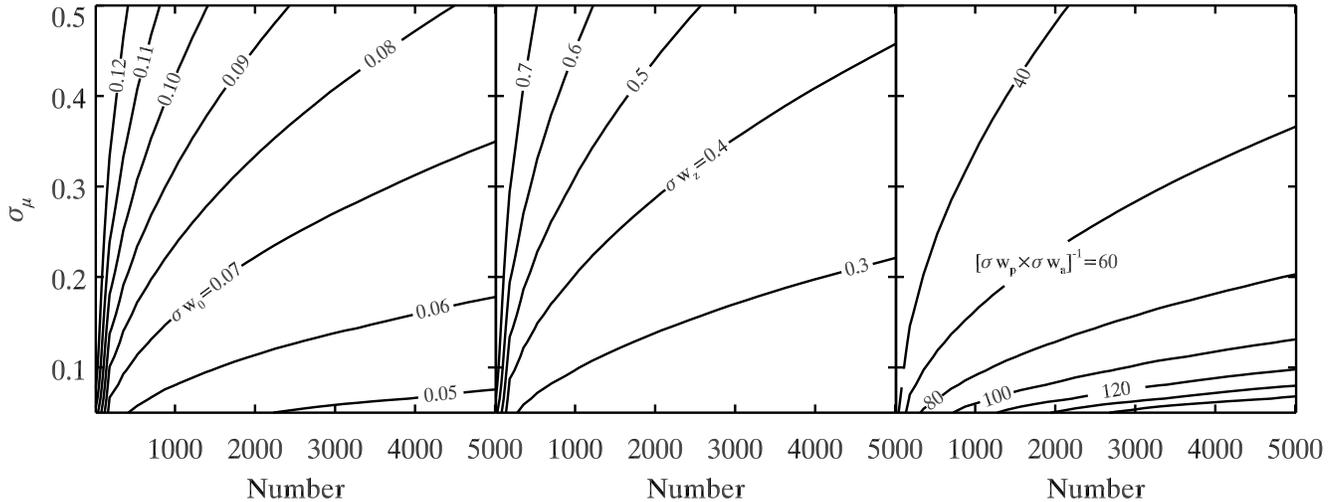}
	\caption{A representation of the trade-off between number of standard candle measurements and the uncertainty in the measurements that dictate the overall dark energy equation-of-state constraints.  Each contour represents a constant  absolute uncertainty in $w_0$ (left) or $w_z$ (middle), or the FoM (right). 
	These constraints are calculated with the Fisher matrix methods using the current SN, CMB, and BAO measurements plus additional $N$ standard candles with a double Gaussian distribution with $(\bar{z}_1,\bar{z}_2,\Sigma_1,\Sigma_2)=(0.0,2.0,0.25,0.25)$.
	The constraints follow the general relationship $\sigma^2_{w_0,w_z} [1/{\rm FoM}] \propto \sigma_\mu^2/N$.}
	\label{fig:sigmanumplot}
\end{figure*}

\subsection{Parameterized Models}\label{sec:linearparamsection}
We consider a CPL dark energy parametrisation, as discussed in Section \ref{sec:parameter}, and initially investigate how the addition of an HzSC affects the likelihood constraints on  a flat $w_z$CDM universe. 

 We combine the constraints from 2000 mock standard candle measurements, uniformly distributed over a large redshift range $0.01\leq z\leq 4.0$, with existing and predicted future SN, BAO, and CMB constraints. The resulting  $\Omega_m-w_0$ and $w_0-w_z$ confidence contours are shown in Fig~\ref{fig:contours}.  With respect to the current constraints, the introduction of the HzSC measurements makes a marked impact on the precision with which we can determine the matter density and equation-of-state parameters. This improvement is mainly attributed to the large number of measurements (despite the lower precision compared to SNe); however, it is also influenced by the large redshift range of the mock standard candle measurements. Having a larger redshift range predominantly reduces the uncertainty in the $w_z$ parameter. However, when the HzSCs are combined with  the predicted future constraints, we find a smaller effect. This is because of the large number and higher precision of predicted future SN data (compared to our mock HzSCs) and the precision and wide redshift range of the predicted future BAO data.
 
 In the $\Omega_m-w_0$ plane, the tilt of the contours is affected by the inclusion of the HzSCs, due to some orthogonality between the probes. As a result, the improvement is more distinguishable than in $w_0-w_z$, where orthogonality is weak.  Consequently, it is difficult to break degeneracies in the equation-of-state parameter using distance probes such as standard candles and standard rulers alone. This degeneracy occurs because $w_0$ does not appear independently in the expression for Hubble parameter. We therefore find that 2000 additional standard candle measurements at high redshift will not improve constraints on this model compared to Stage IV probes, which is not surprising given the relative number of SN and BAO measurements in Stage IV. However, we will see in Section \ref{sec:piecewise} that improvements \textit{are} gained when considering more flexible models of dark energy.

\begin{figure*}
                      \subfloat{ \includegraphics[width=0.45\textwidth]{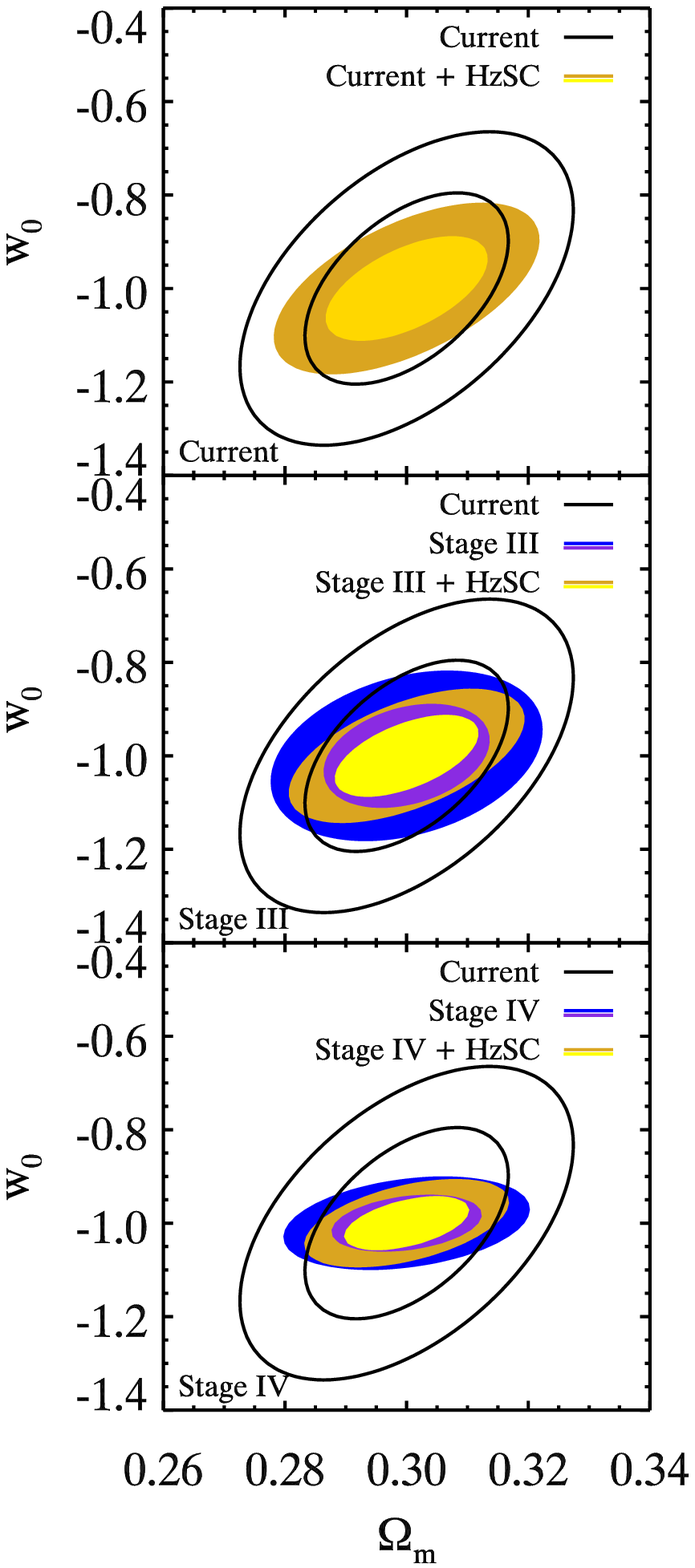}}
                       \subfloat{ \includegraphics[width=0.45\textwidth]{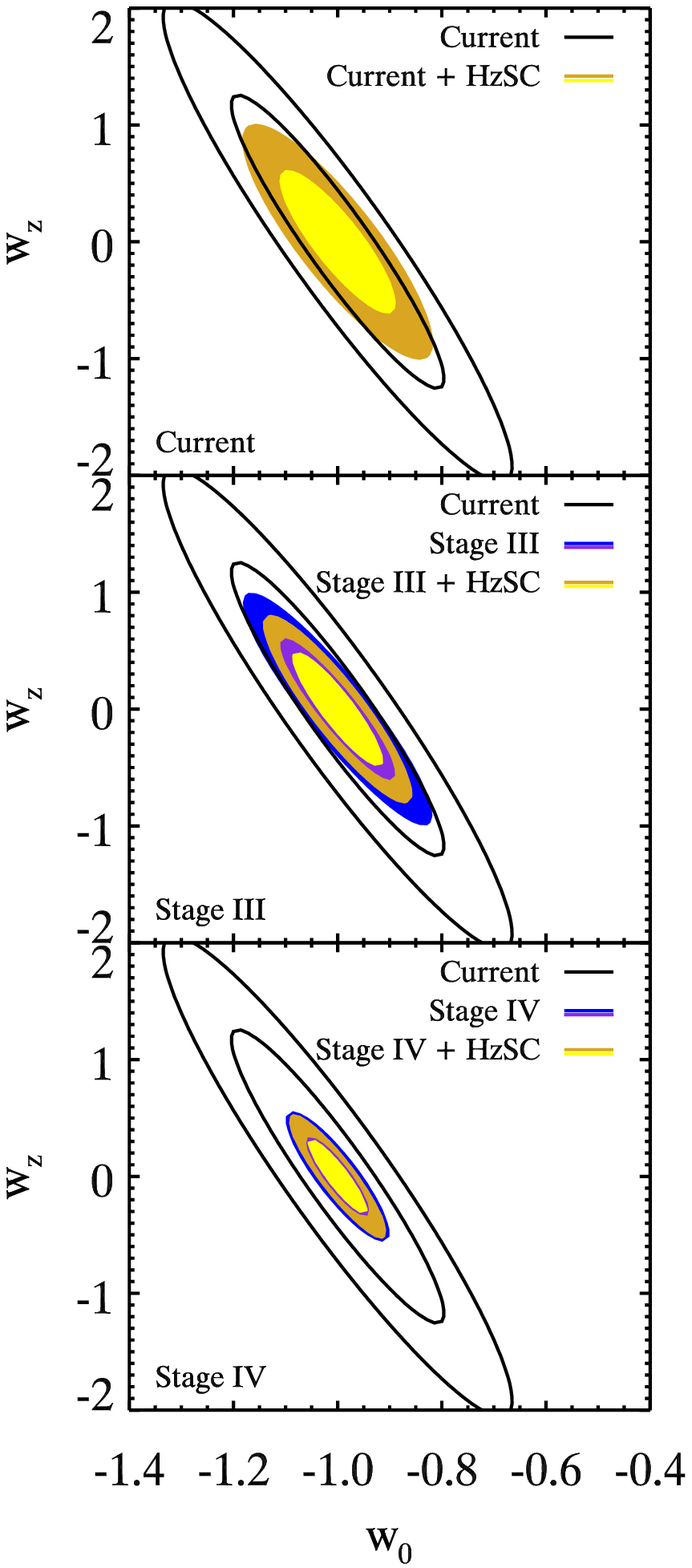}}
                         \caption{The 1$\sigma$ and 2$\sigma$ level confidence contours of the cosmological parameters in the $\Omega_m-w_0$ plane (left) and $w_0-w_z$ plane (right), calculated using Fisher matrix methods. The black contours show the current only SN,  BAO, and CMB constraints, the blue/purple contours show the predicted future constraints from Stage III (middle plot) and Stage IV (bottom plot)  SN, BAO, and CMB data, and the yellow/gold contours in each panel  are the combination of those constraints with HzSCs. HzSCs provide significant improvement over current constraints, and are competitive with Stage III probes.}
                                \label{fig:contours}
\end{figure*}

This initial investigation is not realistic in the sense that it would be infeasible to carry out a HzSC survey with uniform redshift sampling and range for this large number of objects, simply because of the realistic number density and redshift distribution of AGNs, SNe II, or GRBs.  It also gives us very little information about which redshifts are important for dark energy investigations. In order to inform future HzSC surveys of a more optimal, and possibly more realistic, survey strategy, we investigate the constraints on dark energy parameters gained by considering various distributions of HzSCs. To do this, we consider: (1) a uniform redshift distribution and alter its (a) maximum and (b) minimum redshift cut-offs and (2) a redshift distribution described by a Gaussian function, where the mean and range of redshifts probed is varied by altering the mean and width of the Gaussian function.

\subsubsection{Maximum redshift cut-off (Case 1a)}\label{sec:zmax}
We set the minimum redshift of our 2000 HzSC measurements to $\zmin=0.01$  and varied the maximum redshift, $\zmax$, within the range $0.1\leq \zmax\leq4.0$. The HzSC measurements were uniformly distributed in redshift from $\zmin$ to $\zmax$.
For each redshift configuration, we calculate the individual $w_0$ and $w_z$ constraints as well as the FoM. The resulting constraints are shown by the red solid curves in Fig.~\ref{fig:zFoMcombined}. 
In the `Current' case, where the HzSC measurements are combined with the current data, the constraints initially become stronger as the redshift range increases, but there is a maximum in the FoM  at  $\zmax\sim 2$, and beyond this point the constraints weaken. The constraints initially strengthen with redshift as time-evolving behaviour in $w(z)$ becomes easier to identify, and $\Omega_m$ constraints tighten with an extended redshift range.  The turnover is due to a combination of two effects: (1) a uniform HzSC distribution leads to a relative decrease in the number density of low-$z$ measurements (where dark energy is dominant) as the redshift range is extended; (2) by $z\sim2$ the energy density of dark energy in the fiducial model is an order of magnitude smaller than the matter energy density, so its influence on the expansion (and measured luminosity distances) is minimal compared to that of the matter density.  By increasing the $\zmax$ beyond this point, high sensitivity is required to obtain additional information about the dark energy parameters.  Overall, the improvement gained with the addition of the HzSCs to Current cosmological probes is primarily due to the large increase in the total number of distance measurements, but the extended redshift range of the HzSCs also reduces degeneracies between $w_0$ and $w_z$.
The resulting constraints are therefore very dependent on the redshift distribution of HzSCs. It should be noted here that the CPL dark energy parametrization we investigate was expressly designed as a probe of low-$z$ dynamics, so our results are to some extent a consequence of the parametrization choice.

Once Stage III observations have been completed, SNe and BAO will be competitive with the HzSCs we have simulated here. At that point, the orientation of the constraints, in the $w_0$-$w_z$ parameter space, start to play a larger role. Different redshift ranges rotate the degeneracy direction in the $w_0$--$w_z$ plane.
 In this case,  the FoM no longer experiences a turn over and continues to improve with higher $\zmax$ values.  This is a consequence of the other distance probes (SNe and BAO) contributing mostly only at relatively lower redshifts and supports the need for HzSCs, which now complement their lower $z$ counterparts by adding information about the behaviour of dark energy at high $z$.  

By Stage IV, the constraints are already so strong that adding HzSCs gives negligible improvement in the $w_0$ and $w_z$ constraints.  Nonetheless, Fig. \ref{fig:zFoMcombined} shows that the FoM is still improved with the addition of HzSCs at Stage IV because of an increased correlation between the two equation-of-state parameters, thus decreasing the area of the $w_0$--$w_z$ ellipse without significantly reducing its extent in either parameter.  

 Despite the fact that 2000 HzSC measurements, with our prescribed level of measurement uncertainty, only provide a slight improvement on the combined Stage III and Stage IV measurements in the $w_z$ model, we find that 2000 HzSC measurements are overall competitive as individual probes compared to the individual predicted future SN and BAO measurements.
Fig.~\ref{fig:zFoMcombinedcurrentfuture} shows the relative predicted improvement over the current FoM with the individual addition of HzSC measurements, Stage III- and Stage IV- SN and BAO constraints. The HzSC constraints are roughly equivalent to or greater than the predicted Stage III constraints, but to be able to compete with or surpass the Stage IV measurements a large number ($n>2000$) or more precise measurements ($\sigma_{\mu}< 0.2$ mag we assumed here) are still required.

 \begin{figure*}
   \includegraphics[width=1.\textwidth]{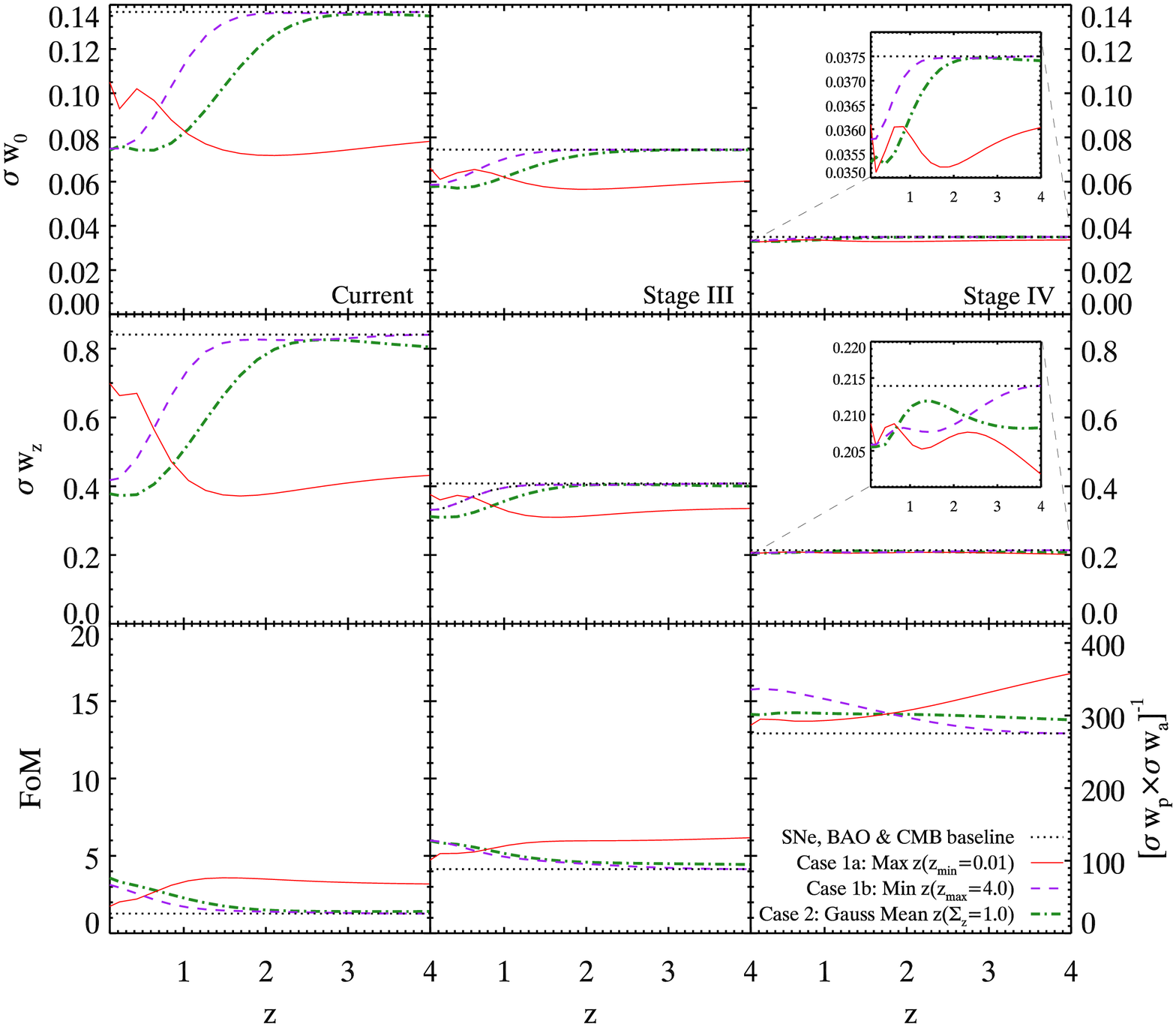}
  \caption{The improvement 2000 standard candles would contribute relative to the dark energy equation-of-state parameter baseline constraints of the combined SN, BAO, and CMB  measurements (black curves) based on current data (left-hand plot) and future Stage III (middle plot) and Stage IV (right-hand plot) data, in a flat $w_z$CDM cosmology. The predicted improvement is shown for the three different HzSC redshift distributions discussed in Section \ref{sec:linearparamsection}. The red (solid) [purple (dashed)] curves show results for Case 1a (1b) where the maximum (minimum) redshift is varied for a uniform distribution of standard candles with a fixed minimum [maximum] redshift at $z=0.01$ ($z=4.0$).  The green (dot-dashed) curves show results for Case 2, where the mean redshift is varied for a Gaussian distribution of standard candles with width $\Sigma_z = 1.0$. Higher FoM values indicate stronger constraints.}
  \label{fig:zFoMcombined}
\end{figure*}

We note also that while we have only shown how HzSCs might improve constraints on $w_0$ and $w_z$ for a flat $w_z$CDM universe model, we also forecasted the effect on other cosmological parameters. The magnitude of the predicted constraints and the behaviour as a function of $\zmax$ depends on the cosmological parameter of interest and the parametrization tested. For example, the density parameters ($\Omega_m$, $\Omega_x$)  always prefer a long redshift range, which is contrary to what is observed in Fig. \ref{fig:zFoMcombined} for the $w_0$ and $w_z$ parameters.  The complexity of the universe model assumed (e.g., $\Lambda$CDM, $w$CDM, $w_z$CDM, or different flavours of $w_z$CDM)  also affects the predicted constraints and changes the optimal redshift distribution. Allowing curvature to vary weakens the strength of the dark energy equations-of-state constraints slightly and tends to shift the optimal $\zmax$ value for $w_0$ and $w_z$ to a lower redshift.  The degradation of the constraints with curvature is expected as there is a  well-known degeneracy between dark energy and curvature for purely geometric probes like standard candles \citep{Linder2005,Knox2006,Huang2007,Hlozek2008}. 

\begin{figure}  
  \includegraphics{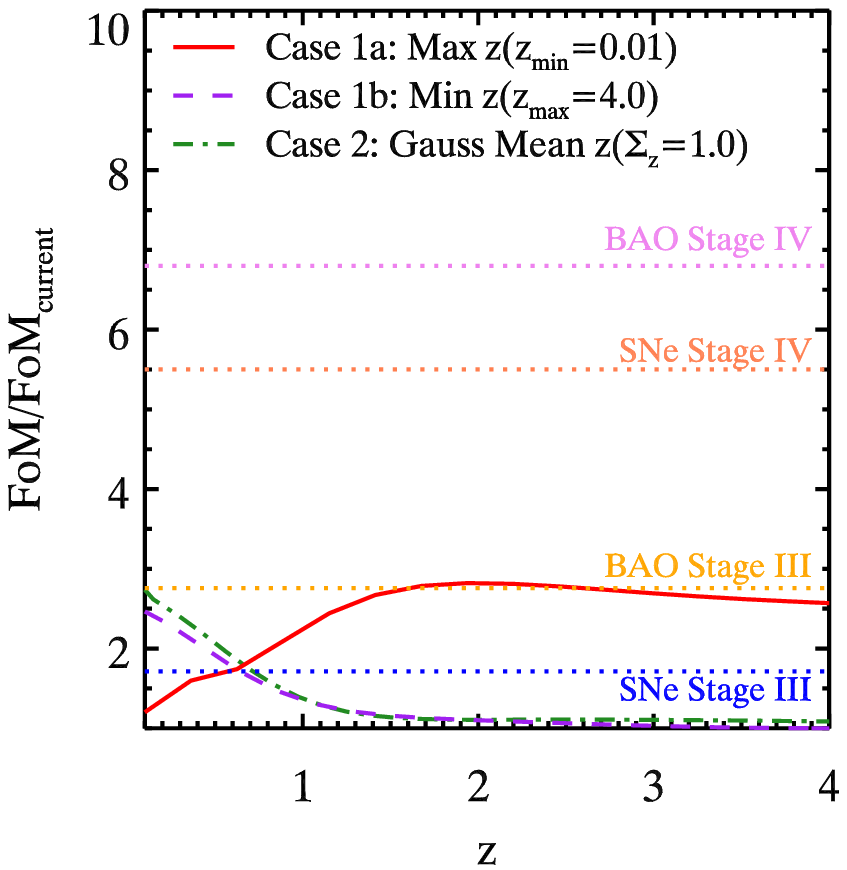}
  \caption{The predicted improvement of the FoM in the $w_0-w_z$ plane in a flat $w_z$CDM model compared to the current constraints.  The non-dotted lines represent the predicted improvement from two uniform distributions with variable maximum (red solid) and minimum (purple dashed) cut-off redshift and one Gaussian distribution with a variable mean redshift (green dot-dashed). These representations are consistent with Fig. \ref{fig:zFoMcombined}. The other lines represent the predicted improvement on the current constraints gained with the different stage future experiments. A 2000 large HzSC survey can marginally compete with the Stage IIIs SNe and BAO but cannot compete with either Stage IV SN or BAO predicted constraints.}
  \label{fig:zFoMcombinedcurrentfuture}
\end{figure}

\subsubsection{Minimum redshift cut-off (Case 1b)}\label{sec:zmin}
Next we consider the effect of changing the minimum redshift. We set the maximum redshift to $\zmax=4.0$ and vary the minimum redshift within the range $0.01\leq \zmin\leq3.9$. 
The purple dashed curves in Fig.~\ref{fig:zFoMcombined} show that all constraints are maximized when the redshift distribution of HzSCs extends to $z=0$. The loss of cosmological information as low-redshift data are removed arises
not only because dark energy is dominant at low redshifts but also due to uncertainties in the Hubble constant  and in the absence of absolute luminosity calibration. In this absence, cosmological information is gained from the shape of the observed Hubble diagram (${\rm d}\mu(z)/{\rm d}z$), rather than the absolute value of $\mu(z)$.  Therefore, long redshift ranges are preferred, and high- and low-redshift standard candle information must be on the same absolute magnitude scale in order to robustly probe the evolution of the expansion and minimize systematic uncertainties as a function of $z$.
As a consequence, in the fiducial cosmology, low-redshift standard candle measurements are just as, or more, important as their high-redshift counterparts.

We further illustrate the dependence of the constraints on the redshift in Fig.~\ref{fig:agnomw0waagnfisherchi} by showing how the standard candle-only likelihood contours, in the $w_0$--$w_z$ plane, change with redshift range. We consider three redshift distributions: 

\begin{enumerate}
\item{$0.01\leq z\leq 1.0$, }
\item{$0.3\leq z\leq 4.0$, }
\item{$0.01\leq z\leq 4.0$.}
\end{enumerate}

When only low redshifts  are probed (case i), the $w_0$ parameter is well constrained, but the $w_z$ parameter is only weakly constrained. At low redshifts, the dynamics of dark energy and consequently the expansion, are predominantly controlled by the value of $w_0$. On the other hand, probing only high redshift information (case ii) may be more sensitive to $w_z$, but without low redshift constraints,  a degeneracy arises between the two equation-of-state parameters and two very different $w(z)$ models  may be hard to distinguish. A visual example of this situation is illustrated in Fig.~\ref{fig:distmods} where different models with curves of the same colour appear almost identical at high redshift.
Instead, it is optimal to probe a broad redshift range (case iii) to gain tight constraints on $w_0$ while restricting the possible values of $w_z$.

Fig.~\ref{fig:agnomw0waagnfisherchi} was created using the $\chi^2$ analysis  instead of the Fisher matrix analysis, as the Fisher matrix analysis is generally not a good representation of the likelihood of individual probes. See Appendix \ref{sec:fishercomp} for more discussion on this point. The presence of the bend or kink in the contours is primarily due to uncertainties in the matter density \citep{Goliath2001,Wolz2012}, which cause degeneracies in the $w_0$ and $w_z$ plane. When standard candle measurements are combined with current CMB and BAO data this uncertainty diminishes significantly, and the bend disappears.
\begin{figure}
		\includegraphics{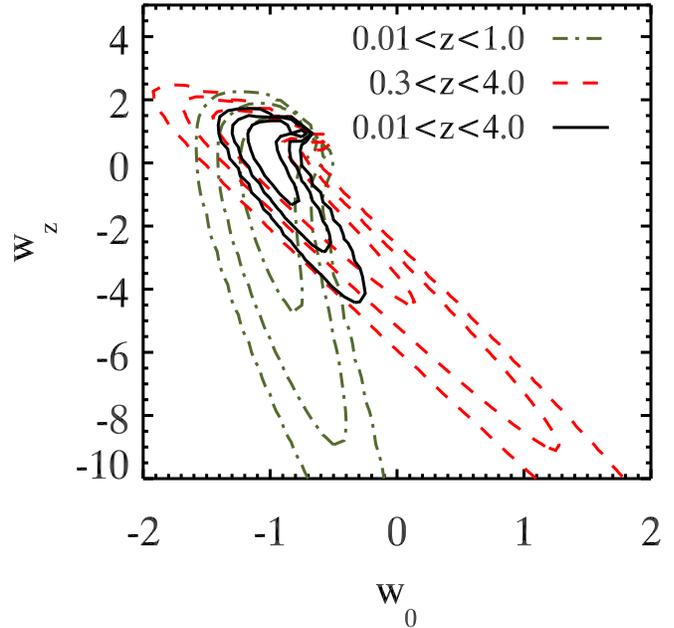}
	\caption{ The 1$\sigma$, 2$\sigma$  and 3$\sigma$ level confidence contours for three different standard candle survey regimes: (a) $0.01\leq z\leq 1.0$ (dot-dash green), (b) $0.3\leq z\leq 4.0$ (dash red), and (c) $0.01\leq z\leq 4.0$ (solid black). Current SN and BAO measurements are restricted to $z\lesssim1$. In this regime, $w_0$ is well constrained, but to constrain $w_z$, we clearly require both high- and low- redshift measurements. The likelihoods were calculated using $\chi^2$ analysis as the HzSC-only constraints are not well approximated as Gaussian, unlike the  constraints obtained from a combination of all the probes.}
		\label{fig:agnomw0waagnfisherchi}
\end{figure}

\subsubsection{Single Gaussian distribution (Case 2)} \label{sec:gauss}
The two previous cases, while instructive, are over-simplified in their assumption of a uniform redshift distribution of standard candles. Indeed, some redshift ranges may have more power in terms of constraining cosmological parameters (see Fig.~\ref{fig:distmods}).  Also, uniform redshift distributions are difficult to achieve in practice due to the small volumes present at low redshift and survey flux limits at larger redshifts even if survey design can attempt to mitigate these effects to some degree.

Here we consider a Gaussian redshift distribution for our mock HzSC sample. 
By varying the mean redshift and width of this Gaussian we can directly probe the relative importance of different redshift regimes in constraining cosmology. We invoke a simple Gaussian distribution with a (variable) mean redshift ($\bar{z}$) and redshift spread ($\Sigma_z$), with the number density of standard candle measurements is given by:
\begin{equation}
\mathcal{N}(z)= \frac{1}{2\pi\Sigma_z^2}e^{-(z-\bar{z})^2/{2\Sigma_z^2}}.
\end{equation}

We initially investigate a constant redshift span by holding the redshift spread $\Sigma_z$ fixed at $\Sigma_z=1.0$.  This is approximately consistent with a relatively deep flux-limited survey (e.g., 2SLAQ; \citet{Croom2009}).  The green dot-dashed curves in  Fig. \ref{fig:zFoMcombined} show the parameter constraints we compute as a function of $\bar{z}$. The distributions are truncated at zero to avoid unrealistic (negative) redshifts, but the total number of HzSC measurements always remains constant.

 In the Current case, the constraints from a Gaussian redshift distribution survey with $\bar{z} =0$, are optimal and the resulting constraints are directly comparable to those found in the uniformly distributed case with redshift within the range $0.01<z<2.0$. This once again solidifies the importance of low-redshift measurements. 
  In the Stage IV case, the FoM constraints gained from a Gaussian distribution of $\Sigma_z=1$ are no longer competitive with the strongest uniform distribution configurations,  and while there is still a preference for a low mean redshift it is not as pronounced as it was in the Current and Stage III cases. The low redshift regime is well constrained by the future SN and BAO measurements, in this case, and broadening the redshift range is now the more efficient approach to constrain dark energy.
 
When the range of redshifts probed, $\Sigma_z$, is also allowed to vary, the constraints do not simply tighten for a decreasing $\bar{z}$ and increasing $\Sigma_z$ as one may naively expect.  Fig.~\ref{fig:gaussandwidth} shows that the optimal value of $\bar{z}$ is marginally dependent on our choice of $\Sigma_z$. For a narrow survey (i.e. small $\Sigma_z$) a small but non-zero $\bar{z}$ is optimal because there are two opposing influences at play: the first, is the preference for low redshift measurements, and the second, is the preference for a larger redshift range.  The relative power of these two preferences depends on the priors imposed.  For the current case, where cosmological constraints are relatively weak,  only an HzSC survey narrower than $\Sigma_z<0.5$ will prefer a non-zero $\bar{z}$, while in the Stage III case, the upper width threshold is closer to $\Sigma_z<0.7$ for a non-zero mean. As our constraints improve with the anticipated Stage III and Stage IV measurements, the preference for a wide redshift range dominates over the need for low redshift measurements.
 
 \paragraph{Double Gaussian:}
 while the global maximum in the FoM shown in Fig.~\ref{fig:gaussandwidth} is at low redshifts it is interesting to note that there is also a small local maximum or plateau in the Current constraints  (shown in the inset of Fig.~\ref{fig:gaussandwidth}).  
 Therefore, information about the dark energy equation-of-state parameter (in the linear parametrization and fiducial model) can be gained from high-redshift data. Due to this, we consider a double Gaussian redshift distribution, $\mathcal{N}(\bar{z}_1,\bar{z}_2,\Sigma_{z(1)},\Sigma_{z(2)})$, which probes both the high- and low-redshift regimes. As low redshift data has been found to be of importance, we set one Gaussian distribution to be stationary at a mean redshift of $\bar{z}_1=0.0$ with a spread in redshift of $\Sigma_{z(1)}=0.25$. 
 We then allow the $\bar{z}_2$ and $\Sigma_{z(2)}$ of the second Gaussian distribution to vary. Each distribution contains 1000 measurements. In Fig.~\ref{fig:doublegauss} the resultant dark energy constraints are shown: in all three cases FoM is maximized when $\Sigma_{z(2)}$ is large. The predicted Current and Stage III  case FoM values are both maximized for $1\leq\bar{z}\leq2$,  while the predicted Stage IV FoM value increases in a roughly monotonic fashion with $\bar{z}_2$.
We find that the constraints predicted based on this redshift distribution are marginally superior to the previous distributions tested (Sections \ref{sec:zmax} and \ref{sec:zmin}). This indicates that the optimal redshift regime for an HzSC for a flat wCDM model is something like a double Gaussian distribution.

It should be noted that having a double Gaussian distribution of HzSC measurements is different from simply using SN data at low redshift and a different standard candle at high redshift, the low- and high-redshift measurements must be calibrated to the same relative distance scale. 

\subsubsection{Conclusion}
In summary, with only our current SN, BAO, and CMB constraints the most efficient way of extracting information about dark energy is to focus our observations on the low-$z$ regime but once the Stage III and IV measurements are completed, and constraints tighten, then it is more informative for HzSCs to probe over a longer redshift range.
  \begin{figure}
   \includegraphics[width=.45\textwidth]{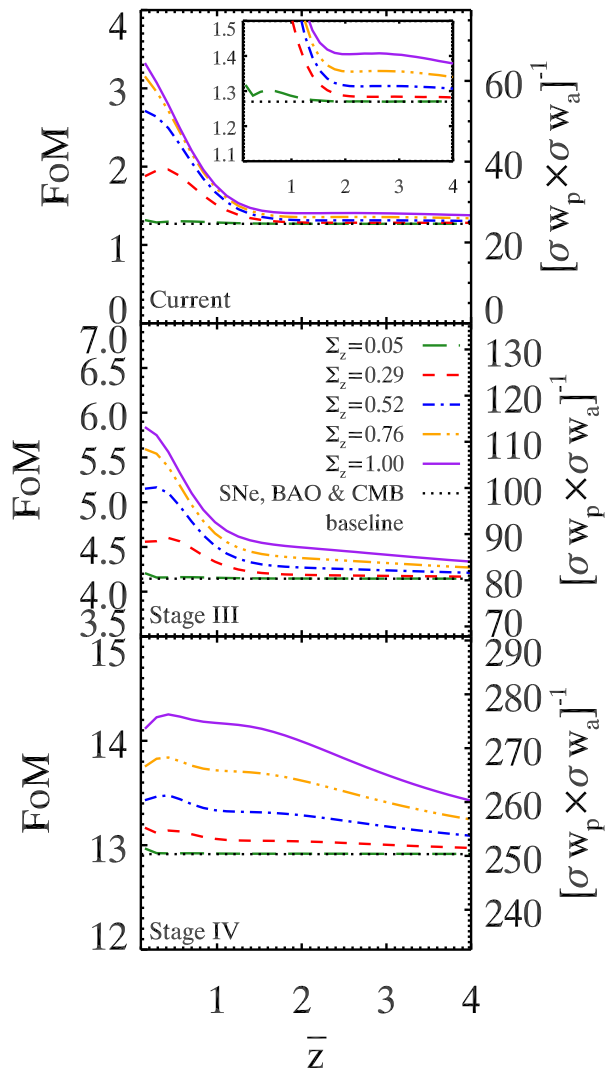}
  \caption{The predicted improvement in the dark energy FoM parameter for 2000 Gaussian distributed HzSCs over the SNe+BAO+CMB baseline (black dotted curves) for Current (top), Stage III (middle), and Stage IV (bottom) constraints, in a $w_z$CDM cosmology. Coloured curves show the predictions as a function of $\bar{z}$ for a single Gaussian distribution of HzSC measurements with different values of $\Sigma_z$. The inset in the top panel shows a zoomed in section of the FoM values.}
  \label{fig:gaussandwidth}
\end{figure}

  \begin{figure}
   \includegraphics[width=.45\textwidth]{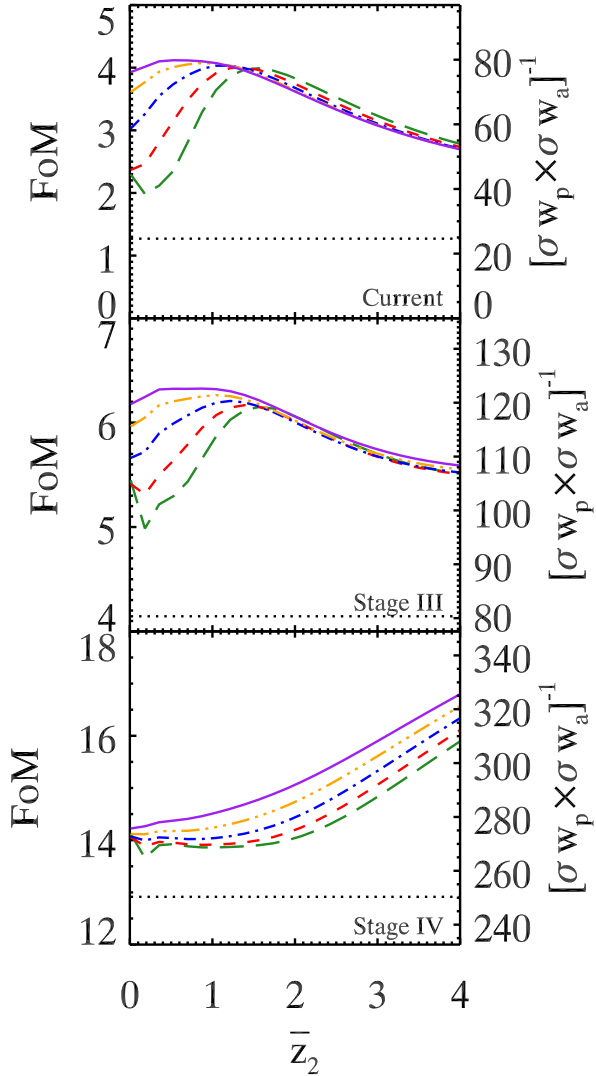}
  \caption{Same as Fig.~\ref{fig:gaussandwidth} except the 2000 HzSC measurements are equally split between a stationary Gaussian distribution with  $\bar{z}_1=0.0$ and $\Sigma_{z(1)}=0.25$, and a Gaussian distribution with variable redshift mean ($\bar{z}_2$) and spread ($\Sigma_{z(2)}$).  Coloured curves represent the same $\Sigma_z$ magnitudes as Fig.~\ref{fig:gaussandwidth} but refer only to changes in $\Sigma_{z(2)}$.}
    \label{fig:doublegauss}
\end{figure}

\subsection{Piecewise models}\label{sec:piecewise}

\begin{figure*}
		\includegraphics[width=1.0\textwidth]{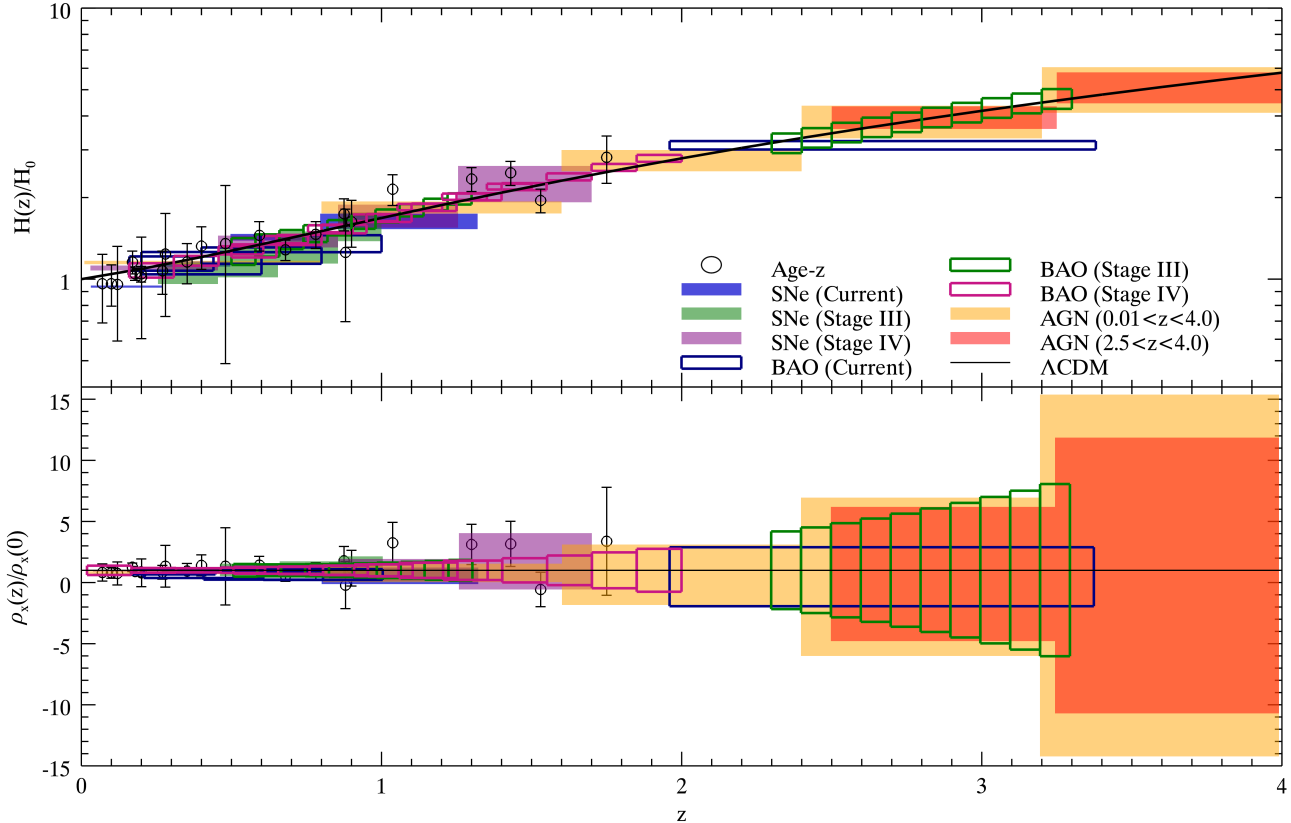}
	\caption{The relative expansion history (Top) and the dark energy density function (Bottom) found in uncorrelated redshift bins from current (blue) and future  (green: Stage III, purple: Stage IV) SN (filled) and BAO (open) data, 2000 uniformly distributed simulated HzSC measurement with two redshift ranges $0.01\leq z\leq 4.0$ (gold-filled) and $2.5\leq z\leq 4.0$ (orange-filled), and other age-$z$ measurements as listed in Table~\ref{tab:hubble} (black-circle).  A fiducial $\Lambda$CDM cosmology $(\om,\oll, w)=(0.26,0.74)$ is shown by the black curve. The current data are consistent with our fiducial model. Note that the boxes for the current BAO measurements enclose the full range of redshifts included in each measurement, but the weighted mean is generally offset from the centre of the bin (e.g. the highest redshift current BAO measurement has an effective redshift of 2.3). High redshift measurements will be able to give constraints in the currently unconstrained redshift regime. }
	\label{fig:hubbleconstantsne}
\end{figure*}

\subsubsection{Hubble Parameter and Dark Energy Density}
We  applied the numerical derivation technique, described in Section~\ref{sec:Hubble}, to the SNLS SN data\footnote{In order to be able to extract the Hubble parameter from the SNLS analysis we set the stretch and colour parameters as constant, with values $\alpha=1.45$ and $\beta=3.16$. These values correspond to the best-fitting values found by \citet{Conley2011} for a flat $w$CDM model when only considering statistical error.},
{future mock SN measurements}, and mock HzSC measurements. The mock HzSC catalogue consists of a uniform distribution of 2000 measurements ranging over $0.01\leq z\leq4.0$. 

The results are shown in the top panel of Fig.~\ref{fig:hubbleconstantsne} with Hubble parameter measurements from existing BAO data \citep{Blake2012,Chuang2012,Busca2012}, existing differential galaxy age measurements \citep{Simon2005,Stern2010,Moresco2012,Zhang2012}, and predicted future BAO data. It should be noted that when we refer to the BAO measurement here, we combine the BAO scale with the Alcock \& Paczynski effect \citep{Alcock1979}. This effect measures redshift space distortions in the shape of the correlation function or power spectrum. It is the combination of these two measurements that gives such precise $H(z)$ measurements.  For more details on the existing Hubble parameter measurements see Appendix \ref{sec:HubbleParamMeas}. AGN or another HzSC and future BAO measurements will probe the high redshift regime. The fiducial 2000 HzSC measurements examined here will not be able to compete with the precision of future high-redshift BAO measurements of $H(z)$, per bin, but they can reach somewhat higher redshifts.

As described in Section~\ref{sec:Hubble}, we can also derive the redshift evolution of the dark energy density, $\rho_x(z)/\rho_x(0)$, assuming a flat universe and given a precise measurement of the matter density fraction. The bottom panel of Fig.~\ref{fig:hubbleconstantsne} shows the resulting estimates for the dark energy density function.  {Current} measurements are consistent with $\rho_x(z)/\rho_x(0)=1.0$ (equivalently, $w=-1$) though the constraints are much weaker in the  high redshift bins.  
 At present, the uncertainty in $\Omega_m$ is a dominant source of uncertainty in the dark energy density function estimation, and due to this, an increase in the precision of the Hubble parameter  will only provide a relatively small improvement in the overall uncertainty levels of the dark energy density function.
Future independent measurements of $\Omega_m$ from CL and lensing will help diminish this restriction.

As the dark energy density function measurements are independent of adjacent redshift bins, and depend only on the local $\partial D_m/\partial z$ derivative, we do not require an absolute luminosity calibration or low redshift measurements to obtain high-redshift information. As a consequence we can concentrate our standard candle measurements in the high-redshift regime. For a $2.5\leq z \leq4.0$ uniform regime, we can increase the precision on the high redshift values of $\rho_x(z)/\rho_x(0)$ measurement by about 25 per cent compared to a $0.01\leq z\leq 4.0$ redshift distribution (orange compared to gold, in Fig.~\ref{fig:hubbleconstantsne}). This is simply due to an increase in the number of measurements in this redshift range. 

The error bars of the dark energy density function depend on both the number of measurements in each redshift bin, as seen above, and the precision of the measurements. Once again, if we assume the main sources of scatter in the standard candle are statistical, rather than systematic, we can consider the improvement in parameter constraints as a trade-off between the number of measurements and the precision of the measurements.  It follows a general $\sigma_{H(z)}^2 \propto \sigma_{\mu}^2/N$ relationship, as observed in the previous section.  Therefore, to compete with the predicted future high-redshift Stage III BAO measurements for the dark energy density function, we require either $\sim20000$ HzSC measurements, with $\sigma_{\mu}=0.2$ mag or, 2000 HzSC measurements, with $\sigma_{\mu}\sim0.06$ mag.

\subsubsection{Direct determination of $w(z)$}
In Section \ref{sec:parameter} we investigated a simple parametrization of the dark energy equation-of-state and found that depending on the parameter and the complexity of the parametrization the optimal redshift distribution was variable. Presently, we do not have a strong theory about the true form of dark energy and to properly investigate the potential of HzSC data, we need to consider a more general form of the dark energy. We adopt a general piecewise step function of the equation-of-state as described in Section \ref{sec:wzpiece} for this purpose, and test how the addition of 2000 uniformly distributed HzSC measurements in the redshift range $0.01\leq z \leq 4.0$ affects the constraints on the dark energy equation of state in combination with existing SN, BAO, and CMB constraints, compared to the  predicted constraints from future surveys.
 
\begin{figure*}
\includegraphics[width=1.0 \textwidth]{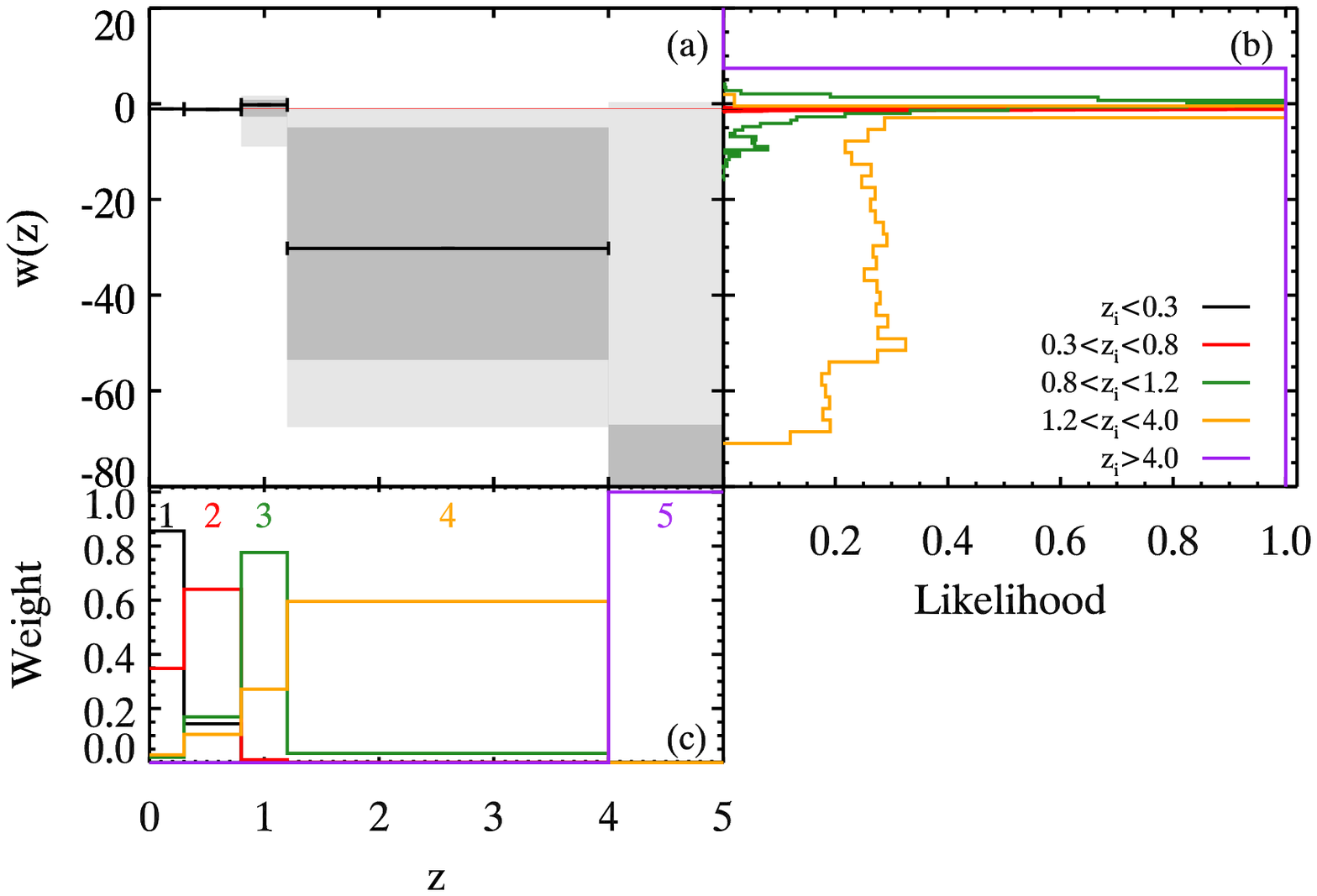}
\caption{Uncorrelated estimates of the derived piecewise dark energy equation-of-state $w(z_i)$ from current SN, BAO, and CMB data (a). The solid black lines correspond to the median value of $w(z_i)$, the dark and light grey shaded regions correspond to the 68 and 95 percent confidence levels and the thin red line corresponds to $w(z)=-1$ ($\Lambda$CDM). This will form the base line in which we add our mock HzSC measurements, Stage III and Stage IV constraints. The coloured histograms in panel (b) show the corresponding normalized likelihood histograms for the five redshift bins we consider. Panel $(c)$ shows the weighting function which transforms the correlated $w(z_i)$ values into uncorrelated $w(z_i)$ values. The new uncorrelated $w_i$ are given as a linear combination of the correlated $w_i$ described by the weight function. There are no prior constraints on the $w_i$ values.}
\label{fig:westcurr}
\end{figure*}

\begin{figure*}
\includegraphics[width=1.0 \textwidth]{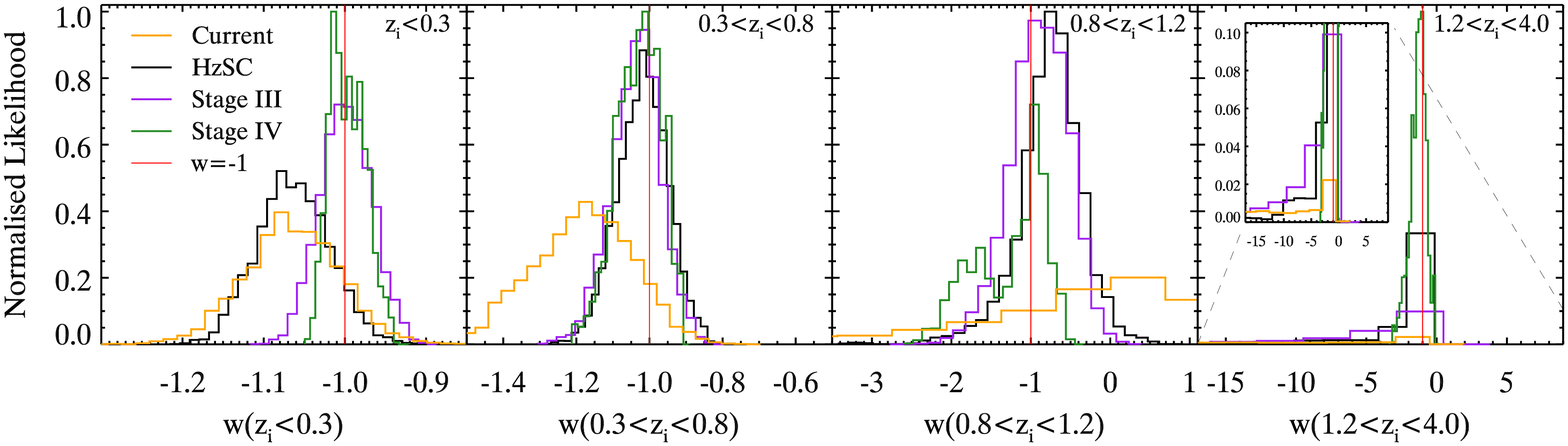}
\caption{The normalized likelihoods of the uncorrelated $w(z_i)$ values for the Current (orange), $0.01<z<4$ HzSC (black), Stage III (purple), and Stage IV  (green) constraints, for the four lowest redshift bins [far left: $z_i<0.3$, middle left: $0.3<z_i<0.8$, middle right: $0.8<z_i<1.2$, far right: $1.2<z_i<4.0$]. The thin red line represents $w(z)=-1$. The inset of the far right plot shows a zoomed in section of the likelihood for $w(1.2<z_i<4.0$). We do not show the $z_i>4.0$ bin because it is poorly constrained.}
\label{fig:likelihoodall}
\end{figure*}

The derived $w(z_i)$ values for the current SN, BAO and CMB measurements are shown in Fig. \ref{fig:westcurr}. This shows the current status of the field and will act as the base line to which we add our mock HzSC measurements, Stage III, and Stage IV constraints.
The lower three redshift bins ($z<1.2$) are well constrained by the current data, but beyond $z=1.2$ the constraints become much weaker. At present, only the CMB measurements and 5 SN measurements contribute to constraining the two highest redshift bins.
The current data are consistent with a cosmological constant ($w(z)=-1$). 
The value of $w(1.2<z_i<4.0)$ is also consistent with $w=-1$ and its maximum likelihood values coincide with  $w=-1$ (see Panel (b) of Fig. \ref{fig:westcurr} and far right plot Fig. \ref{fig:likelihoodall}), but it remains largely unconstrained for values of $w(z)<-1$. The likelihood displays an almost flat distribution tail.  The non-negligible tail in the likelihood curve causes the median value for $w(1.2<z_i<4.0)$ to be significantly offset from the maximum likelihood value. To reduce the extent of  this tail and make strong constraints on the value of $w(1.2<z<4.0)$ we require additional information, for example, HzSC constraints, and/or Stage III and IV constraints. We take a special interest in this redshift range for this reason. Also, despite having the largest redshift range, the last bin is not well constrained, because other parameters, such that $\Omega_m$, $H_0$ and $\Omega_bh^2$ have a more dominant effect than $w(z>4.0)$ on the observed CMB parameters. This remains true for all the cases we consider. 

The weight functions for each redshift bin are shown in Panel (c) of Fig. \ref{fig:westcurr}, where the composition of each redshift bin is distinguished by a different colour (black: $0.0\leq z_1<0.3$, red: $0.3\leq z_2<0.8$, green: $0.8\leq z_3<1.2$, orange: $1.2\leq z_4<4.0$, purple: $4.0\leq z_5$). As an example, the uncorrelated value of $w_i$ for the lowest redshift bin (black) is a linear combination of $w(z<0.3)$ ($\sim85\%$ contribution),  $w(0.3<z<0.8)$ ($\sim14\%$ contribution),  and $w(0.8<z<1.2)$ ($\sim1\%$ contribution). The $ z_5\geq 4.0$ redshift bin (purple) is predominately constrained by CMB measurements (that is, $z>4.0$) and, as a consequence, largely uncorrelated with the lower redshift bins. This is apparent in its weighting function, which has no or little  contribution from the lower redshift bins. In general our low redshift constraints agree with those found by previous authors \citep{Said2013,WangDai2013}, but beyond a redshift $z>0.8$ our constraints were found to be weaker than those found in either \citet{Said2013,WangDai2013}. In both of those studies independent $H(z)$ measurements and a prior on $H_0$ were included in their constraints, and the highest redshift bin was held constant at $w_i=-1$. These factors may explain the discrepancy with our results.

Stage III constraints show a marked improvement in the lower three redshift bins over the current constraints. 
 The $w(1.2<z_i<4.0)$ is also markedly improved, with the introduction of 10 high redshift BAO measurements, however, a long distribution tail is still present. This is evident in the inset of Fig. \ref{fig:likelihoodall} (Purple). As mentioned, the last redshift bin is again not well constrained, and consequently not shown in the histogram, but with the introduction of the Stage III data an upper limit on the value of $w(z_i>4.0)$ becomes apparent. The upper limit appears to be approximately $w(z_i>4.0)<0.7$ (99.99\% upper bound). 
This upper limit arises because a high value of $w$ corresponds to the dark energy behaving more like matter (when $w>-1/3$ dark energy is attractive), and since we have tight constraints on the matter density and Hubble constant, the data can not accommodate more matter at early epochs. Therefore, this upper limit represents the value of $w$ for which the dark energy will have a detectable effect on the CMB measurements.  However, no lower bound is expected as the lower the value of $w$, the more negligible the dark energy density is at early times (recall $\Omega(z) \propto [1 + \Delta z]^{3(1+w)}$)\footnote{This equation only holds true over a redshift range where $w$ remains constant.}, and the more negligible is the dark energy's effect on the expansion.

\begin{figure*} 
\includegraphics[width=1.0 \textwidth]{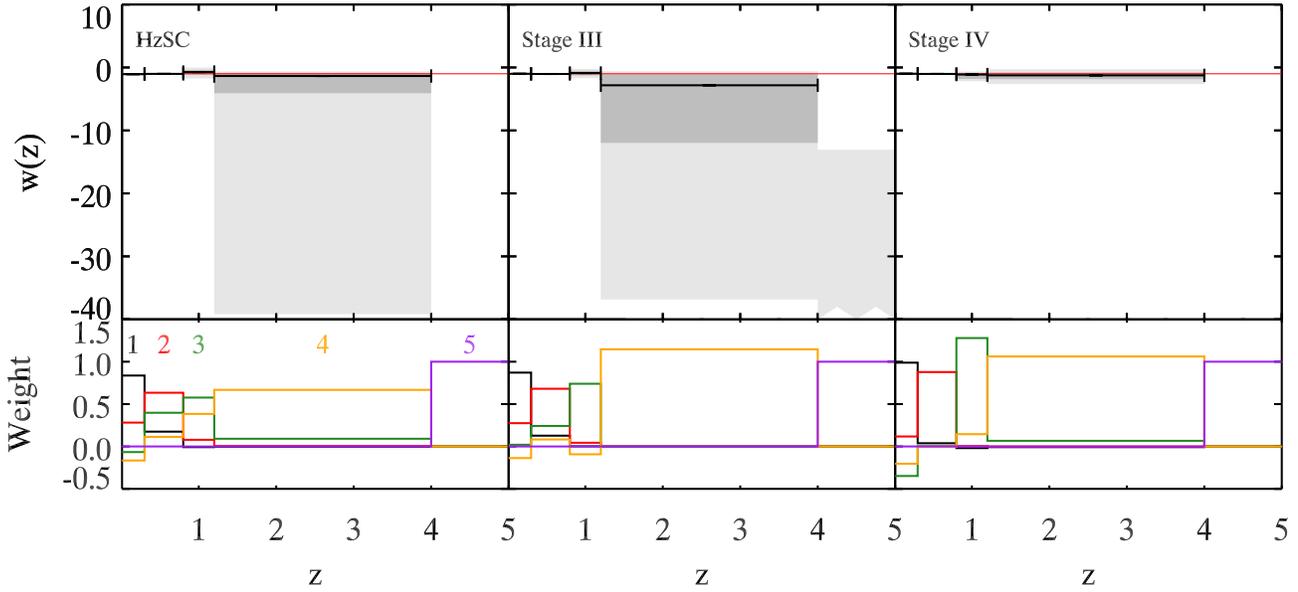}
\caption{Same as Fig. \ref{fig:westcurr} Panel (a) and (c) but for $0.01<z<4.0$ HzSC, Stage III and Stage IV constraints. All these constraints use the current  SN, BAO, and CMB measurements as a baseline.}
\label{fig:westsolid}
\end{figure*}

The addition of HzSC measurements to the current constraints considerably strengthens the  constraints on  $w(1.2<z_i<4.0)$ (Fig. \ref{fig:westsolid}, left). A slight tail in the likelihood distribution is still present (black curve, Fig. \ref{fig:likelihoodall}), but it has a steeper decline relative to the current and Stage III constraints.  As in the Stage III case, an upper limit on $w(z_i>4)$ is observed of approximately $w(z_i>4)\lesssim -0.4$ (99.99\% upper bound). The presence of this predicted upper limit suggests that both Stage III and HzSC constraints may be able to rule out early time dark energy. 

With the introduction of Stage IV constraints, we see significant improvement in all bins (Fig. \ref{fig:westsolid}, right). The Stage IV measurements introduce 530 SN and 27 BAO measurements in the $1.2<z_i<4.0$ bin, distributed according to \citet{Albrecht2006}, and as a consequence can tightly constrain the value of $w(z)$ in this redshift range. It also has the additional advantage of stronger $\Omega_m,~ H_0$, and $\Omega_b$ constraints, which allow the signature of $w(z)$ to be more identifiable.

In the previous section, where we consider a piecewise Hubble parameter and dark energy density function, we introduced a sample of HzSC measurements that only occupied the high redshift regime and found a marked improvement in the high redshift constraints. We have used the same technique here with the piecewise $w(z)$ case, with uniformly distributed HzSC measurements over the  redshift range $1.2<z<4$. The resulting $w(z)$ constraints are shown in Fig. \ref{fig:comboplothigh}. Naively, we would expect to see an improvement in the $w(z_i>4.0)$ constraints as we saw in the Hubble parameter, but due to the reduced redshift range, the constraints on $\Omega_m$ weaken, and correspondingly weaken the $w(z_i>1.2)$ constraints. Therefore, exclusively obtaining high redshift measurements is not beneficial for investigating dark energy, when combined with the current constraints, and a full redshift range is optimal. This is analogous to what we find in the linear $w(z)$ parametrization analysis.
 This did not occur in the dark energy density parameter ($\rho_x(z)/\rho_x(0)$) estimates as the matter density was measured independently.

\begin{figure*}
\includegraphics[width=1.0 \textwidth]{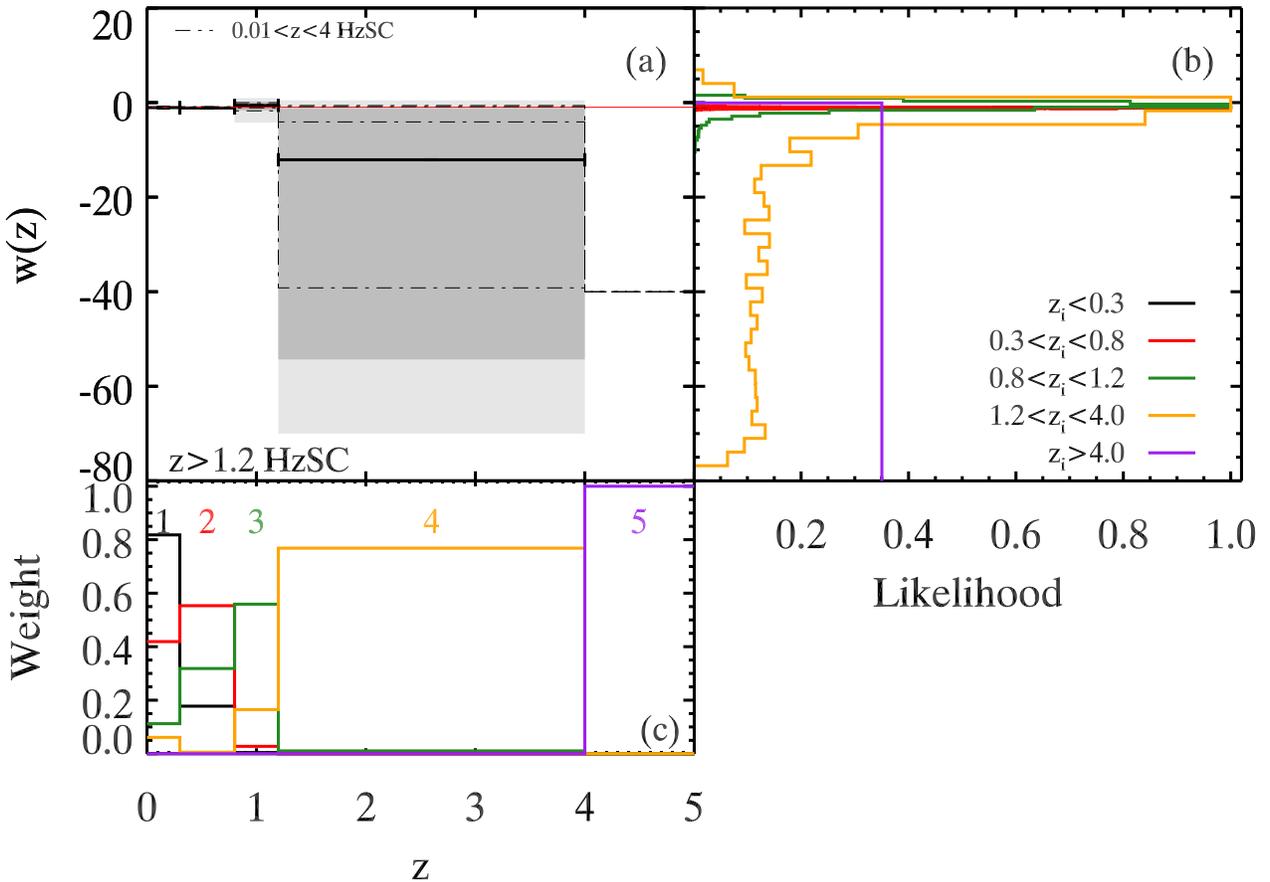}
\caption{Same as Fig. \ref{fig:westcurr} for the combination of current SN, BAO and CMB data with 2000 $z>1.2$ HzSC measurements. In  panel (a) we have overlaid the 68 and 95 percentile constrains for the $0.01<z<4.0$ HzSC constraints (dot-dashed line). }
\label{fig:comboplothigh}
\end{figure*}

Finally, we also considered the addition of 2000 uniformly distributed $0.01<z<4.0$ HzSC measurements in combination with Stage IV constraints. The resulting $w(z_i)$ values and likelihood curves are shown in Fig. \ref{fig:comboplotfuture4agn}. With the addition of the HzSC measurements, the constraints are improved by $\sim30$ per cent. This added precision may give new insight into the nature of dark energy and help to cut down the number of allowable dark energy theories.

As we mentioned earlier, in none of the cases was the last bin well constrained. This bin is constrained solely by the CMB data.
In some previous studies this last bin was held constant at $w=-1$ \citep{Sullivan2007, Sarkar2008,Said2013,WangDai2013}, but we did not want to impose this restriction on our general $w(z)$ expression, as  we did not want our data to  restrict our model and hinder the revelation of surprises in $w(z)$ if they exist. Despite that, if the low redshift bins are consistent with $w(z)=-1$, then the dark energy density becomes negligible at high redshifts. As a consequence, determining  $w(z)$ beyond this point will require a colossal number of precise distance measurements, and as we saw from our predictions, this may only allow us to determine an upper limit. This prediction is highly dependent on our choice of fiducial model. If we have an underlying varying $w(z)$, it may have a large effect on the expansion of the Universe in the high redshift regime and be more easily detectable, but current data do not support this hypothesis. In either case, HzSCs are valuable tools for probing these high-redshift regimes, especially in the presence of exotic forms of dark energy.

\begin{figure*}
\includegraphics[width=1.0 \textwidth]{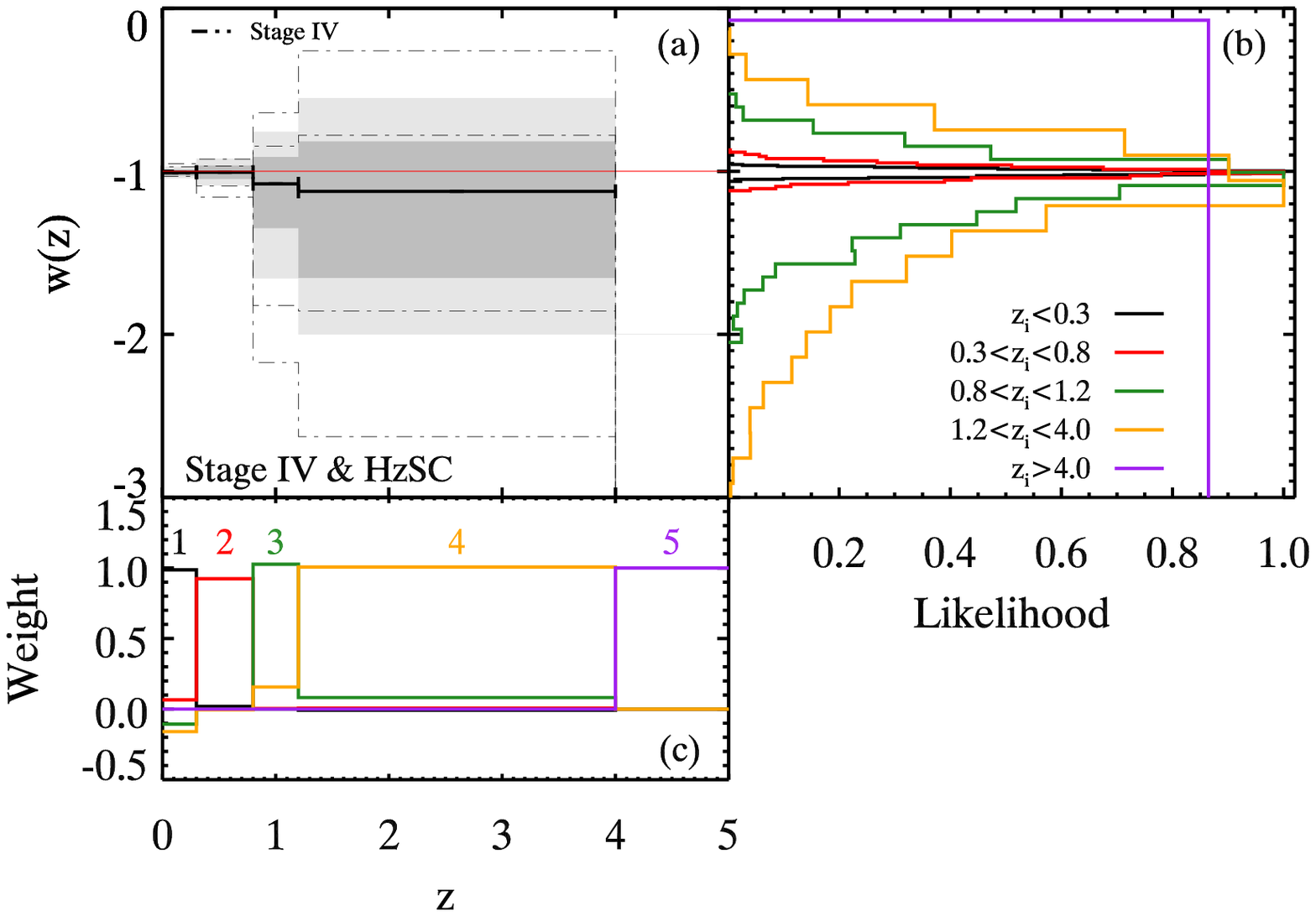}
\caption{Same as Fig. \ref{fig:westcurr} for Stage IV and $0.01<z<4.0$ HzSC constraints. In  panel (a) we have overlaid the 68 and 95 percentile constrains for the Stage IV only constraints (dot-dashed line).}
\label{fig:comboplotfuture4agn}
\end{figure*}

\subsection{Large Scale HzSC Survey}
So far, we have only considered 2000 HzSC measurements. However, if given the same considerations as Stage IV measurements (e.g., time-scale, cost, and researcher hours), including a dedicated telescope and well planned observation strategy, it is not unrealistic to consider of the order of 50,000 HzSCs  with $\sigma_{\mu}=0.2$ or equivalently 12,500 HzSCs  with $\sigma_{\mu}=0.1$.  To investigate the potential of this number of HzSC measurements, we once again consider a flat $w_z$CDM parametrization of the dark energy equation-of-state and uniformly distribute the HzSC measurements over the redshift range $0.01<z<4.0$. The resulting $w_0$--$w_z$ probability contours are shown in Fig. \ref{fig:1e5hzsc}. The constraints calculated for 50,000 HzSCs   measurements (or 12,500 with $\sigma_{\mu}=0.1$) are predicted to be comfortably stronger than the predicted combined Stage IV constraints.

The likelihood of this number is strongly dependent on the observational requirements of the proposed HzSC candidate, and while these predictions do not include any systematic errors which can seriously limit the resulting constraints, HzSCs can have a huge potential for exploring the properties of dark energy given sufficient resources.
\begin{figure}
        \centering
        \includegraphics[width=0.5\textwidth]{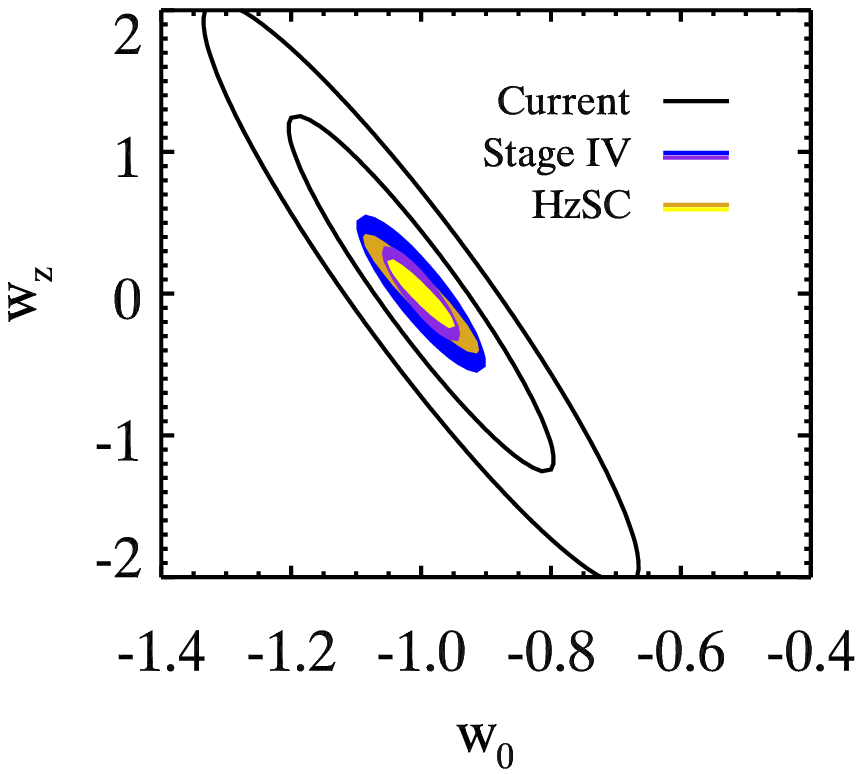}
        \caption{The 1$\sigma$ and 2$\sigma$  level confidence contours in the $w_0$--$w_z$ plane, calculated using the Fisher matrix methods. The black contours show the current SN,  BAO and CMB constraints only, the blue/purple contours show the predicted future constraints from Stage IV SN, BAO and CMB data, and the yellow/gold contours show the constraints from 50,000 uniformly distributed HzSCs with $\sigma_{\mu}=0.2$ or equivalently 12,500 HzSCs  with $\sigma_{\mu}=0.1$, combined with the current constraints. The predicted constraints from 50,000 (12,500) HzSC measurements exceed the predicted constraints from the joint Stage IV SN and BAO measurements.}
        \label{fig:1e5hzsc}
\end{figure}

\section{Discussion and Conclusions}\label{sec:discuss}
Our analysis concentrates on the cosmological constraints that can be obtained from any HzSC. 
We (1) constructed mock standard candle catalogues with a range of distributions, and (2) using Fisher matrix and $\chi^2$ likelihood analyses, predicted how well HzSC measurements could further constrain the dark energy properties when combined with existing and predicted future SN, BAO, and CMB measurements, and (3)  assessed the optimal distribution of HzSC measurements for this type of investigation. Determining whether HzSC measurements could constrain time-evolution in the dark energy equation of state was of primary concern. We approached time-evolution in the equation of state by considering two dark energy models: (1) a linearly parametrized form of the equation-of-state and (2) a piecewise equation-of-state.

The HzSCs show their real strength when constraining a general piecewise $w(z)$ parametrization, especially in the so far unconstrained redshift range, $1.2<z<4.0$. Even with 2000 HzSC measurements, the constraints in that range from an HzSC uniformly distributed with $0.01<z<4.0$ surpass the predicted Stage III constraints, and complement the Stage IV constraints well.  

For the linear parametrization case, we generally found that measurement uncertainty, or equivalently the number, and redshift distribution of the HzSCs both have a large effect on the constraints.  For a uniform distribution of HzSCs, we observed a general strengthening in the $w_0$ and $w_z$ constraints when the maximum redshift was increased, especially when combined with the predicted future SN and BAO constraints. \citet{Linder2003} found analogous results despite using an alternative dark energy model (i.e. $w =w_0 + w'z$). They also argue that having a long redshift baseline decreases the effects of measurement systematics on the $w(z)$ constraints, which we did not consider. Both ours and Linder's results are dependent on the low redshift regime being well constrained. The influence of dark energy on the expansion is greatest at low redshifts, so  in the absence of low-redshift measurements, the constraining power of a standard candle is critically diminished. 
  
 We determined that a double Gaussian-like distribution was optimal for this type of investigation. This agrees with the results of \citet{Frieman2003}, who find the optimal distribution for SN measurements (when combined with CMB measurements) for constraining a linear dark energy equation-of-state is bimodal, with one population at low redshift and the other above a redshift of 1.0 (although they also use a different parametrization of the equation-of-state parameter). \citet{Frieman2003} restrict their investigation to the low-redshift regime, and did not  include BAO measurement prediction. Therefore our work is able to test the optimal regime for an HzSC in the current state of affairs more robustly. 
 
SNe have been predicted to be observable out to $z<2.5$ \citep{Grogin2011,Koekemoer2011}, but the expected number density of SN measurements  beyond $z>1.5$ is low \citep{Albrecht2006, Hook2013}, and the observed scatter is expected to increase with redshift \citep{Albrecht2006,Conley2011}. On the other hand, a large number of AGN (87822 quasars over 3275 deg$^2$) have been measured between a redshift range $0.058<z<5.855$ \citep[SDSS data release 9][]{Paris2012}, making either AGN or an equivalent HzSC, potentially the superior distance probes for achieving the optimal distributions in either the uniform or double Gaussian case, and, therefore constraining $w_z$CDM.

Note that it is not sufficient to combine two different standard candles (for example, strictly HzSC measurements and low-redshift supernovae), unless the two standard candles can be placed on the same relative distance scale.  The advantage of a long lever-arm in redshift is negated if the high- and low-redshift populations are distinct, because two different marginalizations over absolute magnitude are required. We gain cosmological information solely from the overall shape of the Hubble diagram and any uncertainty between the scaling of the high- and low-redshift populations affects the accuracy with which the shape can be reconstructed.

However, if we only consider a single Gaussian distribution, a low-redshift mean was preferable and   in this redshift regime SNe are likely to be the superior probes of $w$ due to the small observational scatter and minimal observational requirements. In saying that, the strongest $w(z)$ constraints gained from a single Gaussian distribution were found to be weaker than the strongest constraints found for either a uniform or double Gaussian distribution.

The technological requirements and observational resources required to use AGN as cosmological distance indicators are already in place. Given access to the necessary resources, it is a real possibility for the community to obtain the prescribed number of AGN observations contemporaneously with Stage III and well before Stage IV is fully completed.

 When combined with the Stage III and Stage IV measurements,  2000 HzSC measurements, with our prescribed level of measurement uncertainty, only provide a slight improvement on the $w_z$CDM constraints. However, we saw in Fig.~\ref{fig:zFoMcombinedcurrentfuture} that 2000 HzSC constraints are competitive with the individual predicted Stage III SN and BAO constraints. Although 2000 HzSC measurements can not compete with the individual Stage IV constraints, a 50,000 large HzSC survey  with $\sigma_{\mu}=0.2$ (or equivalently 12,500 large HzSC survey with $\sigma_{\mu}=0.1$) can obtain significantly superior  $w_0$--$w_z$ constraints than the combined SN and BAO Stage IV constraints (Fig \ref{fig:1e5hzsc}). This number is highly optimistic, but depending on the observational requirements of the proposed HzSC, it may be possible to accomplish within the Stage IV timeline. Additionally, a measurement uncertainty of $\sigma_{\mu}=0.1$ is feasible to achieve for AGN \citep[][ Kilerci-Eser et al. 2014, in preparation]{Watson2011}.

Regardless of the ability for the HzSC to independently constrain dark energy, it will nonetheless act as an independent probe with respect to all other methods, thereby providing a means to more effectively intercalibrate and evaluate systematic uncertainties across the different methods.  This is not only a desired but a critical aspect of cosmological distance measurements if we are to constrain the dark energy equation of state.

Interestingly, the amplitude of the measurement uncertainties not only directly affect the constraints that an HzSC can place on all the cosmological parameters,  where smaller uncertainties provide stronger constraints. However, they also influence the choice of `optimal redshift distribution' of the HzSC that can place the best constraints on the dark energy equation-of-state parameters. For example, when considering a double Gaussian distribution, once again anchoring one low-redshift Gaussian with $\bar{z}_1=0.0$ and $\Sigma_{z(1)}=0.25$, and allowing the mean redshift of the second Gaussian to vary, we find that as the scatter is reduced the constraints tighten, as expected, but we also find that the optimal mean redshift tends towards a lower value. 
 This suggests that we could tailor our observation strategy to the quality of the probe.

\paragraph*{Caveats:}
generally, we only consider a flat universe, and though flatness is well supported by observations \citep{Planck2013}, this assumption may influence the resulting constraints. We briefly considered the effect of loosening this assumption,  and  the effects varied depending on the model and parameter of interest.  The behaviour of the equation-of-state parameter constraints were not hugely affected by the choice of flatness, though the optimal redshift range was altered slightly.

Throughout our investigations we assumed a $\Lambda$CDM model as our fiducial model to construct our mock catalogues of future measurements. As a consequence, the constraints we derive are affected by this choice. Also we do not consider any source of systematic errors in our predictions, but systematic errors can become a dominant limitation in dark energy investigations. We have not included systematic effects in this investigation, as presently we have insufficient knowledge of the candidate HzSCs to predict the type and magnitude of the possible systematics that may arise. As a consequence, the results of this study are limited by this omission, and represent the most optimistic case. It is crucial for future studies, in which possibly AGN, GRBs, or core collapse SNe are measured, that the systematics are fully investigated. 

 Gravitational lensing magnification, due to intervening structure along the line of sight, introduces scatter in our luminosity distance measurements and will have a degrading effect on the associated constraints \citep{Holz2005}.
However, rather than just being a source of noise, that adds scatter to the magnitudes, the scatter actually has a specific signature, and measuring the lensing signal in standard candles is an interesting new way to test theories of dark energy, because it measures the effect of the distribution of mass along the line of sight on the paths of the photons \citep{Smith2014}. Therefore some of the information lost due to the increase in scatter may be gained from studying the lensing signal.

 It should be noted that we do not consider modified gravity models in our investigations, so whether high redshift candles are useful tools in modified gravity models depends on how far the  models' predicted expansion history deviates from a flat-$\Lambda$CDM model. Other types of measurements, such as growth of structure and cosmic shear, will likely be very useful in distinguishing between such models and standard $\Lambda$CDM \citep{Cardone2012}.   Coincidently, AGNs can also provide extra insight into gravity theory, as  reverberation mapping can directly measure the mass of the central supermassive black hole \citep{Peterson2004, Bentz2009, Denney2010, Grier2012,Barth2013}.

\paragraph*{Conclusion:}
HzSCs can be useful in constraining cosmological models, particularly those with a temporally varying dark energy.  The number and accuracy needed for standard candle measurements to be competitive with future high-redshift SN and BAO measurements will require  significant, long-term observing programs. Nonetheless, seeking to obtain HzSC measurements is a worthwhile enterprise considering (1) the nature of dark energy remains unknown, so gaining additional understanding of it is a significant priority, and (2) there are added benefits of obtaining independent and complementary cosmic distance measurements as a means to further inter-calibrate and cross-check current methods. Also, by using AGN as our HzSC, we will help shed light on their nature and on galaxy -- black hole  co-evolution, since using reverberation mapping methods doubles the use of these measurements, allowing for the mass of the quasar black hole to be measured as well.
 In a forthcoming paper, we investigate the requirements of a realistic, competitive AGN survey more closely. 

\section*{Acknowledgements}
AK would like to
acknowledge the support provided by the University of Queensland and Australian Commonwealth Government via the Australian Postgraduate Award.
TMD acknowledges the support of the Australian Research Council through a Future Fellowship, FT100100595, and the ARC Centre of Excellence for All Sky Astrophysics, CE110001020.
MV acknowledges support from a
FREJA Fellowship granted by the Dean of the Faculty of Natural Sciences at the
University of Copenhagen and a Marie Curie International Incoming Fellowship.
MV thanks the Kavli Institute for Theoretical Physics at University of California,
Santa Barbara for their hospitality while finalizing this work.
The research leading to these results has received funding from the People
Programme (Marie Curie Actions) of the European Union's Seventh Framework
Programme FP7/2007-2013/ under REA grant agreement No. 300553 (MV and KD).
KD would like to acknowledge support from the National Science Foundation through grant no. AST-1302093.
The Dark Cosmology Centre is funded by the Danish National Research Foundation. This research was supported in part by the National Science Foundation
through Grant No. NSF PHY11-25915 to the Kavli Institute for Theoretical Physics (MV).

\bibliographystyle{mn2efix}
\bibliography{agnbib}

\appendix
\section{Analytical marginalization of constant}\label{sec:fishermarg}
If an observable has the form, $x = f(\mathcal{P})+K$, where $K$ is a constant, and if no prior knowledge of $K$ is assumed at all,  the general $\chi^2$ function (Equation \ref{eq:chi2}) can be integrated analytically over ($K\in [-\infty,\infty]$). The resulting revised $\bar{\chi}^2 $ equation is then given by the expression \citep{Goliath2001,Conley2011},
\begin{equation}
\bar{\chi}^2 = A-\frac{B^2}{C},
\label{eq:revisechi2}
\end{equation}
where $A$ is equivalent to original $\chi^2$ equation
\begin{equation}
A= \sum_{ij} \left[x_i^m(\mathcal{P}_{\rm mod})-x_i^d\right] C_{ij}^{-1}\left[ x_j^m(\mathcal{P}_{\rm mod})-x_j^d\right],
\end{equation}
and
\[
B= \sum_{i} \left[x_i^m(\mathcal{P}_{\rm mod})-x_i^d\right] C_{ii}^{-1},
\]
\[
C = \sum_{ij} C^{-1}_{ii}.
\]

\section{Mathematics for Fisher Matrix calculations}\label{sec:fishercalc}
The Fisher matrix is formally defined as the expectation value of the derivatives of the log of the likelihood with respect to the parameter $\lambda$,
\begin{equation}
F_{\alpha\beta} = -\left\langle \frac{\partial^2 \ln \mathcal{L}}{\partial \lambda_{\alpha}\partial \lambda_{\beta}}\right\rangle,
\end{equation}
or more simply the second derivative of $\chi^2$ centred on the best-fitting value. This comes from the Taylor expansion of $\chi^2$ (corresponding to the likelihood function) around the best-fitting value.  Because the best-fitting value corresponds to a minimum in $\chi^2$ (i.e. $d\chi^2/d\lambda|_{\lambda_0} = 0$),  the second order term, $d^2\chi^2/d\lambda^2|_{\lambda_0}$, becomes the most important term. Comparing these two definitions of the Fisher matrix, we can recover the general $\chi^2$ equation, which includes the correlation coefficient, $\rho$, 

\begin{equation}
\chi^2 = \frac{\left(\frac{\Delta x}{\sigma_x}\right)^2+\left(\frac{\Delta y}{\sigma_y}\right)^2 - 2\rho \left(\frac{\Delta x}{\sigma_x}\right)\left(\frac{\Delta y}{\sigma_y}\right)}{1-\rho^2}.
\label{eq:chicorr}
\end{equation}

For the analysis of SNe and HzSCs, our two observables are $z$ and $\mu$, and the error in redshift is negligible, so the elements of the Fisher matrix for this analysis are
\begin{equation}
F_{\alpha\beta} = \sum_z \frac{1}{\sigma_{\mu}^2} \frac{\partial \mu(z)}{\partial \lambda_{\alpha}}\frac{\partial \mu(z)}{\partial \lambda_{\beta}}.
\label{eq:specfisher}
\end{equation}
We define a new parameter $\mathcal{E} = H^2(z)/H_0^2 =E^2(z)$, such that 

\begin{eqnarray}
\frac{\partial \mathcal{E}}{\partial \Omega_m} =&~(1+z)^3 -b(z),\\
\frac{\partial \mathcal{E}}{\partial \Omega_{x}} =&~ f(z)- b(z),\\
\frac{\partial \mathcal{E}}{\partial w} =&~ \Omega_x \frac{\partial f(z)}{\partial w}.
\end{eqnarray}
where $b(z) = (1+z)^2$, which corresponds to the curvature term and is included, as $\Omega_k$ is dependent on the values of $\Omega_m$ and $\Omega_{x}$, such that $\Omega_k = 1-\Omega_m-\Omega_{x}$. Also $f(z) = (1+z)^{3(1+w_0+w_z)}e^{-3w_zz/(1+z)}$.
For the linear (CPL) parameterisation of dark energy, $w(z) = w_0+w_zz/(1+z)$,
\begin{eqnarray}
\frac{\partial \mathcal{E}}{\partial w_0} =&~ \Omega_x \frac{\partial f(z)}{\partial w_0}\\
=&~ 3\Omega_x f(z) \ln (1+z),
\end{eqnarray}
and
\begin{eqnarray}
\frac{\partial \mathcal{E}}{\partial w_z} =& \Omega_x \frac{\partial f(z)}{\partial w_z}\\
=& 3\Omega_x f(z) \left(\ln (1+z)-\frac{z}{1+z}\right),
\end{eqnarray}
 Redefining the comoving distance as $\chi' =\chi(H_0R^{-1}c^{-1})$ for simplicity,
\begin{equation}
\frac{\partial \chi'}{\partial \lambda_i} = -\frac{1}{2}\int_0^z \frac{1}{E^3(z')}\frac{\partial \mathcal{E}(z')}{\partial \lambda_i} dz'~~ {\rm for}~ \lambda_i\in(\Omega_m,\Omega_{x},w_0,w_z).
\end{equation}
The  dimensionless tangential comoving distance, $D_M'=(H_0/c)D_M$, can then be expressed as
\begin{equation}
D_M' =\frac{1}{\sqrt{|\Omega_k|}}S_k \left(\sqrt{|\Omega_k|} \chi'\right)= \frac{1}{\sqrt{\Omega_k}}\sinh \left(\sqrt{\Omega_k} \chi'\right),
\end{equation}
therefore
\begin{eqnarray}
\frac{\partial D_M'}{\partial \lambda_i} = -\frac{1}{2}\frac{1}{\Omega_k^{3/2}} \frac{\partial \Omega_k}{\partial \lambda_i}\sinh \left(\sqrt{\Omega_k} \chi'\right)\nonumber\\ 
 + \frac{1}{\sqrt{\Omega_k}}\cosh \left(\sqrt{\Omega_k} \chi'\right) \left(\frac{1}{2}\frac{1}{\sqrt{\Omega_k}} \frac{\partial \Omega_k}{\partial \lambda_i} \chi' + \sqrt{\Omega_k} \frac{\partial \chi'}{\partial \lambda_i}\right)
\end{eqnarray}
for $w_0$ and $w_z$ parameters ${\partial \Omega_k}/{\partial \lambda_i}=0$. Therefore
\begin{equation}
\frac{\partial D_M'}{\partial \lambda_i} =  \cosh \left(\sqrt{\Omega_k} \chi'\right)  \frac{\partial \chi'}{\partial \lambda_i}.
\end{equation}
Since we are assuming a flat universe in our investigations, we can use the Taylor expansions
\[ \cosh(x) =1 +x^2/2+\ldots\]
\[\sinh(x) = x+x^3/6+\ldots\]
therefore
\begin{equation}
\lim_{\Omega_k \to 0}\frac{\partial D_M'}{\partial \lambda_i} =  \frac{\partial \chi'}{\partial \lambda_i}~~ {\rm for}~ \lambda_i\in(w_0,w_z).
\end{equation}
For $\Omega_m$ and $\Omega_x$ parameters ${\partial \Omega_k}/{\partial \lambda_i}=-1$, therefore,
\begin{eqnarray}
\frac{\partial D_M'}{\partial \lambda_i} = \frac{1}{2}\frac{1}{\Omega_k^{3/2}} \sinh \left(\sqrt{\Omega_k} \chi'\right) + \frac{1}{\sqrt{\Omega_k}}\cosh \left(\sqrt{\Omega_k} \chi'\right)\nonumber\\ 
\times\left(-\frac{1}{2}\frac{1}{\sqrt{\Omega_k}}  \chi' + \sqrt{\Omega_k} \frac{\partial \chi'}{\partial \lambda_i}\right).
\end{eqnarray}
Once again substituting in the Taylor expansion
\begin{equation}
\lim_{\Omega_k \to 0}\frac{\partial D_M'}{\partial \lambda_i} =  \frac{\partial \chi'}{\partial \lambda_i}-\frac{\chi'^3}{6}~~ {\rm for}~ \lambda_i\in(\Omega_m,\Omega_{x}).
\end{equation}

The solutions above were given in \citet{Bassett2011}, but are shown here for completeness.

\subsection{Derivatives of Observables}
For the analysis of SNe and HzSCs, we measure the distance modulus, so 
\begin{equation}
\frac{\partial \mu}{\partial \lambda_i} = \frac{5}{D_L\ln (10)} \frac{\partial D_L}{\partial \lambda_i}= \frac{5}{D_M' \ln (10)} \frac{\partial D_M'}{\partial \lambda_i}
\end{equation}
It is simple to then substitute this into Equation~\ref{eq:genfisher}.

For BAO, we consider the parameters $d_z$ and $A$. Therefore the derivatives of importance are
\begin{equation}
\frac{\partial A}{\partial \lambda_i} = \frac{\sqrt{\Omega_m}}{z}\frac{\partial D_V'}{\partial \lambda_i} + \frac{D_V'}{2z\sqrt{\Omega_m}}\frac{\partial \Omega_m}{\partial \lambda_i},
\end{equation}
where $D_V' = (H_0/c)D_V$ and
\begin{equation}
\frac{\partial D_V}{\partial \lambda_i} = \frac{D_V'}{3}\left(-\frac{1}{2E(z)^2}\frac{\partial \mathcal{E}(z)}{\partial \lambda_i}+\frac{2}{D_M'}\frac{\partial D_M'}{\partial \lambda_i}\right),
\end{equation}
and also
\begin{equation}
\frac{\partial d_z}{\partial \lambda_i} = \frac{1}{D_V}\frac{\partial r_s}{\partial \lambda_i}-\frac{r_s}{D_V^2}\frac{\partial D_V}{\partial \lambda_i}+\frac{\partial D_V}{\partial z_d}\frac{\partial z_d}{\partial \lambda_i},
\end{equation}
where 
\begin{equation}
\frac{\partial r_s}{\partial \lambda_i} = -\frac{c}{2\sqrt{3}H_0}\int_0^{a} \frac{da}{a^2\sqrt{1+(3\Omega_b/4\Omega_{\gamma})a}} \frac{1}{E^3} \frac{\partial \mathcal{E}}{\partial \lambda_i}.
\end{equation}

For the CMB, we consider the parameters $\ell_A$, $\mathcal{R}$, and $z^{*}$.

\begin{equation}
\frac{\partial \mathcal{R}}{\partial \lambda_{i}} = \sqrt{\Omega_m}\frac{\partial D_M'}{\partial\lambda_i}+ \frac{D_M'}{2\sqrt{\Omega_m}}\frac{\partial \Omega_m}{\partial \lambda_i}+  \frac{\partial r_s}{\partial z^{*}}\frac{\partial z^{*}}{\partial \lambda_i}
\end{equation}

\begin{equation}
\frac{\partial \ell_A}{\partial \lambda_i} =\pi\left(-\frac{D_M'}{r_s(z_{*})^2}\frac{\partial r_s(z_{*})}{\partial\lambda} + \frac{1}{r_s(z_{*})}\frac{\partial D_M'}{\partial\lambda} + \frac{\partial \ell_A}{\partial z^{*}}\frac{\partial z^{*}}{\partial \lambda_i}\right)
\end{equation}

We have omitted the explicit derivatives for ${\partial \ell_A}/{\partial z^{*}}$, ${\partial z^{*}}/{\partial \lambda_i}$, as they are trivial to calculate. The Fisher Matrix method is only used to determine the dark energy equation-of-state parameters constraints for the linear (CPL) parametrization.

\section{Supernova Fitting Procedure and Colour-Stretch Correction}\label{sec:Snestuff}
SN measurements have an added complexity of a `colour-stretch correction' in the determination of their magnitude calculation. The expected magnitude of the SNe is then taken to be
\begin{equation}
m_B = 5\log_{10} D_L(\mathcal{P},z_{\rm cmb},z_{\rm hel})-\alpha(s-1)+\beta\mathcal{C}+M_B
\end{equation}
where $z_{\rm cmb}$ is the CMB frame redshift, $z_{\rm hel}$ is the heliocentric redshift, $s$ is the stretch parameter, $\mathcal{C}$ is the colour parameter, and $\alpha$ and $\beta$ parametrize the $s$-luminosity and $\mathcal{C}$-luminosity relationships. In this work, we substitute $\mu$ with $m_B$ for the supernova analysis.

The SNLS uses a combination of two light curve fitting software packages, SALT2 and SiFTO, to determine the stretch and colour parameters. They give equal weight to the fit parameters determined from each of the two software packages and include the difference between the results from each package in their systematic uncertainty budget. To fit the data in $w(z)$ analysis we have marginalized over the stretch and colour parameters. The low-mass ($M_{\rm stellar}\leq 10^{10}M_{\odot}$) and high-mass ($M_{\rm stellar}> 10^{10}M_{\odot}$) host galaxy populations are fitted separately, as prescribed by \citet{Conley2011}. In the Hubble parameter and dark energy density function fitting we set the colour and stretch parameters to constant values of $\alpha=1.45$ and  $\beta=3.16$. These values correspond to the best-fitting values found by \citet{Conley2011} for a flat $w$CDM model when only considering statistical error. Incorrect fitting of the stretch and colour parameters can cause cosmological discrepancies \citep{Conley2011}. 
 
  The covariance matrix is a combination of a systematics covariance matrix, and two covariance matrices which contain statistical errors in the SN model used in the light-curve fit. This follows the procedure outlined in \citet{Conley2011}.
  
\section{Hubble Parameter Measurements}\label{sec:HubbleParamMeas}
Existing Hubble parameter measurements from various sources \citep{Simon2005,Stern2010,Blake2012,Busca2012,Chuang2012,Moresco2012,Zhang2012} are given in Table~\ref{tab:hubble}. These measurements use different probes. The majority of the measurements come from relative age measurements of galaxies at different redshifts, using a variety of techniques, to infer the Hubble parameter \citep{Simon2005, Stern2010, Moresco2012, Zhang2012}. The relative age measurements are used as an estimator for $dz/dt$, which in turn gives $H(z)$.
 \citet{Simon2005} use observations of passively evolving galaxies and synthetic stellar population models to constrain the age of the oldest stars in the galaxy. \citet{Moresco2012} consider the 4000\AA\ break (D4000) as a function of redshift and use stellar population synthesis models to theoretically calibrate the dependence of the differential age evolution on the differential D4000. \citet{Zhang2012} look at the evolution of luminous red galaxies (LRG), while \citet{Stern2010} consider how red-envelope galaxies evolve with time. The Hubble parameter estimates from  \citet{Chuang2012}, \citet{Blake2012}, and \citet{Busca2012} are determined from BAO scale measurements, sometimes in conjunction with other measurements, such as an Alcock-Paczynski measurement \citep{Blake2012} or CMB measurements \citep{Busca2012}.

\begin{table*}

  \caption{Hubble parameter versus redshift from various sources and their corresponding analysis techniques, as described in Appendix~\ref{sec:HubbleParamMeas}.  }
    \begin{tabular}{lllll}
    \hline
    $z$     & $H(z)$  & $\sigma_{H}$ & Method & Reference \\
          & (km s$^{-1}$ Mpc$^{-1}$) & (km s$^{-1}$ Mpc$^{-1}$) &       &  \\
    \hline
    0.07  & 69    & 19.6  & LRG-Age-$z$ & \citet{Zhang2012} \\
    0.1   & 69    & 12    & Age-$z$+SNe & \citet{Simon2005} \\
    0.12  & 68.6  & 26.2  & LRG-Age-$z$ & \citet{Zhang2012} \\
    0.17  & 83    & 8     & Age-$z$+SNe & \citet{Simon2005} \\
    0.179 & 75    & 4     & D4000-Age-$z$ & \citet{Moresco2012} \\
    0.199 & 75    & 5     & D4000-Age-$z$ & \citet{Moresco2012} \\
    0.2   & 72.9  & 29.6  & LRG-Age-$z$ & \citet{Zhang2012} \\
    0.27  & 77    & 14    & Age-$z$+SNe & \citet{Simon2005} \\
    0.28  & 88.8  & 36.6  & LRG-Age-$z$ & \citet{Zhang2012} \\
    0.35  & 82.1  & 5     & BAO   & \citet{Chuang2012} \\
    0.352 & 83    & 14    & D4000-Age-$z$ & \citet{Moresco2012} \\
    0.4   & 95    & 17    & Age-$z$+SNe & \citet{Simon2005} \\
    0.44  & 82.6  & 7.8   & BAO +AP & \citet{Blake2012} \\
    0.48  & 97    & 62    & Red Envelope Galaxies: Age-$z$ & \citet{Stern2010} \\
    0.593 & 104   & 13    & D4000-Age-$z$ & \citet{Moresco2012} \\
    0.6   & 87.9  & 6.1   & BAO +AP & \citet{Blake2012} \\
    0.68  & 92    & 8     & D4000-Age-$z$ & \citet{Moresco2012} \\
    0.73  & 97.3  & 7     & BAO +AP & \citet{Blake2012} \\
    0.781 & 105   & 12    & D400-Age-$z$ & \citet{Moresco2012} \\
    0.875 & 125   & 17    & D4000-Age-$z$ & \citet{Moresco2012} \\
    0.88  & 90    & 40    & Red Envelope Galaxies: Age-$z$ & \citet{Stern2010} \\
    0.9   & 117   & 23    & Age-$z$+SNe & \citet{Simon2005} \\
    1.037 & 154   & 20    & D4000-Age-$z$ & \citet{Moresco2012} \\
    1.3   & 168   & 17    & Age-$z$+SNe & \citet{Simon2005} \\
    1.43  & 177   & 18    & Age-$z$+SNe & \citet{Simon2005} \\
    1.53  & 140   & 14    & Age-$z$+SNe & \citet{Simon2005} \\
    1.75  & 202   & 40    & Age-$z$+SNe& \citet{Simon2005} \\
    2.3   & 224   & 8     & BAO+WMAP7 & \citet{Busca2012} \\
    \hline
    \end{tabular}%
  \label{tab:hubble}%
\end{table*}%


  \section{Fisher matrix method}\label{sec:fishercomp}
The Fisher matrix analysis is a very popular method of predicting the capabilities of future surveys, and we have used this method throughout our analysis for computational simplicity. Despite its popularity the Fisher Matrix method has come under some criticism in previous studies \citep{Wolz2012,Khedekar2012}, as it  only considers Gaussian errors in the parameter space, and therefore can not accurately estimate asymmetric likelihood distributions.  This means that Fisher analyses can reach `forbidden' regions, and cannot completely trace degeneracies in the data that exist in reality. Its applicability also depends strongly on the stability of the derivatives of the likelihood and parameters. If the derivatives are unstable near the chosen model, the Fisher matrix will not be able to accurately represent this behaviour. However, when the constraints from all the probes are combined, the $w_0-w_z$ contours are reasonably Gaussian and the Fisher matrix method can be used to  predict the constraints on these parameters quite accurately. When we test the validity of the Fisher matrix method compared to the $\chi^2$ analysis, we generally found an agreement between the two methods to within 10 percent.
\end{document}